\documentclass[sigconf]{acmart}



\copyrightyear{2026}\acmYear{2026}
\setcopyright{cc}
\setcctype{by}
\acmConference[PACT '26]{International Conference on Parallel Architectures and Compilation Techniques}{October 19-22, 2026}{Chicago, IL, USA}
\acmBooktitle{International Conference on Parallel Architectures and Compilation Techniques (PACT '26), October 19-22, 2026, Chicago, IL, USA}
\acmDOI{10.1145/3838684.3846872}
\acmISBN{979-8-4007-2915-7/2026/10}

\AtBeginDocument{%
  }

\usepackage{def}
\usepackage{booktabs}
\usepackage{algorithm}
\usepackage{algpseudocode}
\usepackage{xspace}
\usepackage{enumitem}

\begin{document}

\title{\pname{}: Unlocking Fine-Grained GPU Concurrency in Vision-Language-Action Models}

\author{Anna Li}
\affiliation{%
  \institution{University of Toronto}
  \country{}
}

\author{Christina Giannoula}
\affiliation{%
  \institution{Max Planck Institute for Software Systems (MPI-SWS)}
  \country{}
  }

\author{Nandita Vijaykumar}
\affiliation{%
  \institution{University of Toronto}
  \country{}
}






\renewcommand{\shortauthors}{Trovato et al.}


\begin{abstract}
Vision-Language-Action (VLA) models have emerged as foundational models for next-generation robotics. 
High VLA inference throughput is critical for meeting the control-rate requirements of robots.
VLA models comprise two phases, a vision-language model (VLM) and an action head, that can be decoupled and executed asynchronously and concurrently across independent robot requests. Through a detailed characterization of four state-of-the-art VLAs, we observe that GPUs are severely underutilized in VLA inference as the Cooperative Thread Array (CTA) scheduler of GPU is unable to fully overlap the two independent phases of VLA execution. We identify that this inefficiency is caused by head-of-line blocking in the hardware thread-block dispatcher which causes the action head phase kernels to be stalled behind large VLM kernels despite idle resources being available. 

We demonstrate that prior scheduling frameworks do not address the challenges posed by VLA concurrency. First, the independent phases across different robot requests each comprise numerous kernels, and at any given time, there are different combinations of kernels that are executed in parallel. This makes static or ahead-of-time scheduling policies largely ineffective. Second, many of the action-head operators are short-running kernels and there are numerous such kernels. This leaves no headroom for online profiling or preemption-based mechanisms. To address these challenges, we present \pname{}, a lightweight GPU scheduling framework that leverages online Streaming Multiprocessor (SM) utilization and individual kernel resource requirements to intelligently and dynamically co-schedule kernels to efficiently overlap the two phases of execution by (1) mitigating head-of-line blocking, and (2) co-scheduling kernels with complementary resource requirements. We demonstrate in simulation, across two GPU architectures, for 4 state-of-the-art VLA models, that \pname{} delivers throughput gains of up to $28\%$, with an average gain of $20.5\%$. 

\end{abstract}

\maketitle

\pagestyle{plain}

\section{Introduction}
 
Embodied systems such as robotic arms, humanoids, and autonomous vehicles are increasingly used in real-world environments; however, they are only useful when they can operate beyond controlled lab environments and handle the variability of real-world settings. The recent breakthroughs in foundational Vision-Language Models (VLMs)~\cite{paligemma, qwenvl, prismatic} tackle this constraint: VLMs contain broad visual and semantic priors from internet-scale image-text data, encoding world knowledge beyond what robot datasets cover. Vision-Language-Action (VLA) models extend the VLM with an action head fine-tuned on robot trajectories. VLAs take as input a video stream and language instructions, and translate latent representations into end-to-end motor commands~\cite{cogact, pi, nvgr001, openvla}. By inheriting the VLM's knowledge, VLAs generalize to objects, scenes, and instructions beyond their robot training distribution, and have therefore become the foundation for next-generation robotics.

VLA models translate natural-language instructions and a stream of frames into low-level motor actions that robots execute. This is achieved through a two-phase inference process: a VLM phase encodes the inputs into an intermediate representation, which an action expert then decodes into executable motor commands as output. While robotic platforms typically require control frequencies near or above 50 Hz~\cite{openvla_oft, tinyvla, smolvla}, VLA inference latencies may not meet these requirements on edge devices and single GPUs~\cite{openvla, efficient_vla_survey, vlaperf}. 
High VLA inference throughput is critical since the action generation rate affects how quickly the robot can process new observations and react to dynamic environments~\cite{dynamicvla}. 
Given limited inference latency and throughput, VLA models have adopted action chunking, where the VLA emits a longer sequence of actions for each inference process. However, during this window, the robot cannot react to new observations, producing abrupt motions~\cite{rtc, adaptive_chunking}.

Recent VLA architectures decouple VLM execution from the action head~\cite{pi, nvgr001} so that the two phases can be pipelined across frames. This is enabled by two techniques: predicting a short sequence of future actions from a single VLM output rather than only the next action~\cite{rtc, faster}, and running the two phases on separate execution streams~\cite{duocore, fis}. Together, these allow the action head for frame $N$ to execute concurrently with the VLM for frame $N+1$, pipelining the two-phase execution~\cite{duocore, smolvla}.

We conduct a comprehensive characterization of four production VLA models on modern GPU architectures and make three key observations. 
First, VLAs significantly underutilize the GPU during inference: median Streaming Multiprocessor (SM) throughput across all four models ranges from only 18\% to 30\% of peak, and even at its busiest, the action phase leaves Tensor Cores below 3\% utilized. 
Second, the two phases occupy different regions of the roofline: vision encoding and VLM prefill have high arithmetic intensity and average 37.4\% Tensor Core utilization due to large GEMMs. In contrast, the action expert operates with a batch size of 1 and contains many low-intensity kernels. This creates an opportunity to overlap Tensor-Core-dominant compute-bound kernels with CUDA-Core execution of bandwidth-bound kernels.
Third, concurrently executing the VLM and action phases causes near-zero slowdown for the VLM phase. However, pipelining does not translate into throughput gains. Naively dispatching the two phases on separate CUDA streams yields a counterintuitive result: the action head experiences significant slowdown, causing action chunks to be emitted at roughly half the rate of the action head running alone, despite negligible interference from the VLM phase and underutilized memory bandwidth.

We observe that this inefficiency arises from the hardware thread-block (CTA) scheduler's head-of-line blocking behavior~\cite{gilman_sigmetrics, kitsune, gpupool, paella}. Because the VLM launches massive grids that exceed the GPU's spatial capacity, the dispatcher saturates all SMs almost exclusively with VLM thread blocks. Despite ample headroom in CUDA Cores on every SM throughout the VLM's execution, the hardware scheduler serializes the two phases, leaving these execution units idle.

Resolving this inefficiency and underutilization presents two challenges specific to VLA workloads: 

\textbf{Challenge 1: Non-Deterministic Kernel Co-Residency.} The VLM kernel co-resident with each action-head kernel shifts at microsecond granularity. Because the two streams progress at independent rates, the VLM kernel executing when a given action-head kernel is dispatched varies across frames. 

\textbf{Challenge 2: Microsecond Kernel Granularity.} The action head is dominated
by hundreds of micro-kernels with a median duration of a few microseconds, even after optimizations and kernel fusion by backends such as TorchInductor (via \texttt{torch.compile}).

Together, these challenges render existing scheduling approaches ineffective for VLAs. Preemption-based hardware schedulers such as Maestro~\cite{Maestro} and SMK~\cite{smk} incur context-switching overheads that eclipse micro-kernel execution times. Online-profiling schedulers such as Warped-Slicer~\cite{warpslicer} require sampling windows that exceed the micro-kernel duration. Software-based approaches~\cite{ispa, aker, tacker, kitsune} require source-code access, making them incompatible with the closed-source vendor libraries pervasive in VLAs. They also demand re-compilation whenever the action head or kernel implementations change. Runtime interposition layers~\cite{tally} depend on specific launch sequences and cannot react in time.

Our key insight is that the information needed to co-schedule kernels from both phases is \emph{static per kernel}: a kernel's utilization of Tensor Cores, ALU pipelines, and memory bandwidth can be measured once offline. The thread-block dispatcher can then dispatch kernels with \emph{different} resource utilization to the same SM. This approach requires no kernel modification and is insensitive to kernel duration, making it robust to micro-kernel granularity.

We propose \pname{}, a GPU scheduling framework that extends the thread-block dispatch path in three ways to support efficient concurrent execution of multiple streams.
First, we extend the kernel launch descriptor with three 4-bit fields $(\tau_{TC}, \tau_{ALU}, \tau_{MEM})$, which encode the fraction of peak Tensor Core throughput, ALU throughput, and DRAM bandwidth that a kernel's CTAs consume. These tags are obtained via a one-time offline profiling pass using vendor counters \cite{ncu} and injected at launch time by a lightweight runtime wrapper \cite{cudaapi}. 
Second, we augment the work distributor's per-SM resource accounting with three 8-bit utilization counters $(U_{TC}, U_{ALU}, U_{MEM})$ that are updated on the GPU's CTA admission path and completion using the kernel's pre-profiled utilization values, leveraging the dispatcher's de-allocation signals. 
Third, we replace the dispatcher's greedy admission policy with a two-phase admission process that commits at most one CTA per admission decision. Phase 1 is responsible for admitting at least one CTA from every active kernel with pending work to prevent micro-kernel starvation behind monolithic GEMMs, and Phase 2 fills the remaining SM capacity by selecting the kernel whose intensity vector best aligns with the SM's headroom. If there is only one pending kernel, then that kernel will simply be selected. 

We implement \pname{} on Accel-Sim~\cite{accelsim} configured for the RTX 4090 and RTX 5090, using NVBit traces of four representative VLAs: CogACT~\cite{cogact}, $\pi_{0.5}$~\cite{pi}, Octo~\cite{octo}, and GraspVLA~\cite{graspvla}. We demonstrate that \pname{} delivers throughput gains of up to $28\%$, with an average gain of $20.5\%$ on both RTX 4090 and RTX 5090.

This paper makes the following contributions:
\begin{itemize}[leftmargin=*, topsep=2pt, itemsep=2pt, parsep=0pt]
    \item We perform a detailed analysis of GPU utilization on state-of-the-art Vision-Language-Action (VLA) models and find that head-of-line blocking in the CTA dispatcher significantly limits execution concurrency of the two phases of VLA inference.
    \item We propose a lightweight scheduling framework that modifies the GPU hardware scheduler with a new admission algorithm and per-SM state-registers to address this gap and demonstrate its effectiveness on four state-of-the-art VLA models.
    \item We demonstrate that across 4 state-of-the-art VLA models, \pname{} delivers throughput gains of up to 28\%, with an average gain of 20.5\%. 
\end{itemize}

\vspace{-1em}
\section{Background and Motivation}

\subsection{Vision-Language-Action Models}
\label{sec:vla}

Traditional robots excel at repetitive tasks in structured settings but struggle in unstructured, real-world environments~\cite{pi, kawaharazuka_survey}. The Vision-Language-Action (VLA) model is an emerging alternative to traditional robotics. A VLA model couples a pretrained Vision-Language Model (VLM)~\cite{llama2, paligemma, qwenvl, prismatic} with a dedicated action head fine-tuned on domain-specific robotic datasets~\cite{bridge, openx, droid}. The VLM acts as a visual-semantic encoder, leveraging internet-scale information to process observations and natural language instructions as inputs. The action head is trained on robot-trajectory datasets and maps the high-level representations generated by the VLM into low-level, continuous motor commands, such as 7-degree-of-freedom (7-DoF) end-effector controls, for the robot to execute. VLA models are an emerging and promising technique for embodied AI~\cite{kawaharazuka_survey, pure_vla_survey}. However, because action generation relies on computationally expensive VLMs, VLA inference latency currently fails to meet the high-frequency closed-loop requirements of modern robotics. Hence, improving the inference efficiency of VLA models has become a critical challenge in modern robotics. 

\begin{figure}[!htbp] 
    \centering
    \includegraphics[width=\linewidth]{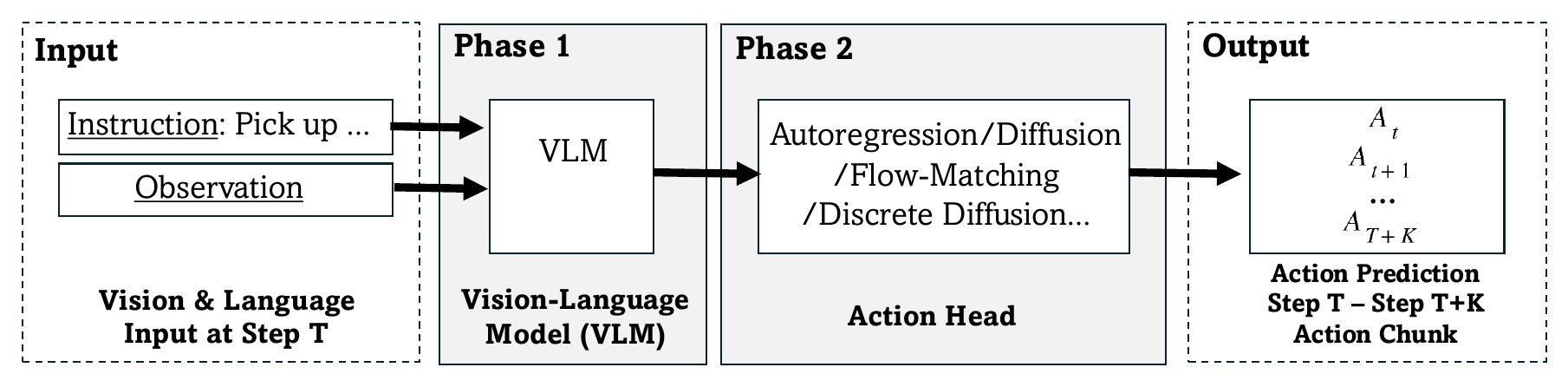}
    \vspace{-0.6cm}
    \caption{Representative VLA architecture decomposed into two phases. 
    }
    \label{fig:vla-arch}
\end{figure}

Figure~\ref{fig:vla-arch} illustrates a representative VLA architecture. In Phase 1, the model takes as input a frame and a textual instruction, propagating them through a VLM to generate an intermediate representation. In Phase 2, this representation is passed to the action head to predict a series of future actions, known as an \emph{action chunk}, to be executed consecutively by the robotic arm. The action head produces these robotic actions through a variety of mechanisms, including autoregressive token generation \cite{openvla, rt2, openvla_oft, fast, spatialvla, univla}, diffusion action heads \cite{diffusion_policy, octo, cogact, rdt1b, hybridvla}, flow-matching action heads \cite{pi, nvgr001, smolvla, gr3}, and most recently, discrete diffusion action decoders \cite{ddvla, dvla, mmadavla}.

\vspace{-1em}

Early VLA models relied on autoregressive token prediction~\cite{rt1, rt2}, with OpenVLA~\cite{openvla} remaining the dominant open-source baseline. Subsequent works replaced discrete token prediction with continuous-action generation. Diffusion Policy~\cite{diffusion_policy} laid the foundation for generating robotic actions via a denoising process, and was adopted in Octo~\cite{octo}, RDT-1B~\cite{rdt1b}, and CogACT~\cite{cogact}.
Recently, flow matching has emerged as the dominant action-head architecture. Pioneered by $\pi0$~\cite{pi0}, flow-matching now serves as the action head driving both industrial models ($\pi0.5$~\cite{pi}, GR00T N1/N1.5~\cite{nvgr001}, GR-3~\cite{gr3}, Gemini Robotics 1.5~\cite{geminirobotics}) and open-source models (SmolVLA~\cite{smolvla}, AsyncVLA~\cite{asyncvla}). Emerging in late 2025, discrete diffusion as an action-head paradigm takes two distinct forms. One branch eliminates the traditional action head in favor of a monolithic, end-to-end transformer~\cite{ddvla, mmadavla}, while the other retains the action-head abstraction but uses discrete denoising instead of continuous decoding~\cite{dvla, lladavla, udvla}.

Despite the rapid emergence of various VLA architectures, flow matching remains the dominant paradigm. We therefore focus on flow-matching-based VLAs in this paper, along with diffusion-head VLAs, which share the same two-phase structure and resource-utilization profile. We will focus on and evaluate four representative open-source models: \textbf{CogACT}~\cite{cogact}, a diffusion-head model with a 300M DiT action head; \textbf{Octo}~\cite{octo}, a lightweight diffusion-head model with a compact 3M action head; \textbf{$\pi$0.5}~\cite{pi}, an industrial flow-matching model representative of the frontier deployment regime; and \textbf{GraspVLA}~\cite{graspvla}, a flow-matching model with a Chain-of-Thought perception-then-action pipeline. Together, these models span diffusion action heads at two parameter scales \cite{cogact, octo} and flow-matching action heads from both industry~\cite{pi} and academia~\cite{graspvla}. All four follow the common two-phase structure of Figure~\ref{fig:vla-arch}: phase 1 (VLM phase) encodes images and language and produces an intermediate state consumed by phase~2 (action-head phase), which generates motor commands. There are slight differences in how phase 1's output is encoded: as readout-token embeddings \cite{octo}, cognition features \cite{cogact}, KV cache \cite{pi} or generated token sequences \cite{graspvla}.


\begin{figure}[!htbp]
    \centering
    \includegraphics[width=1\linewidth]{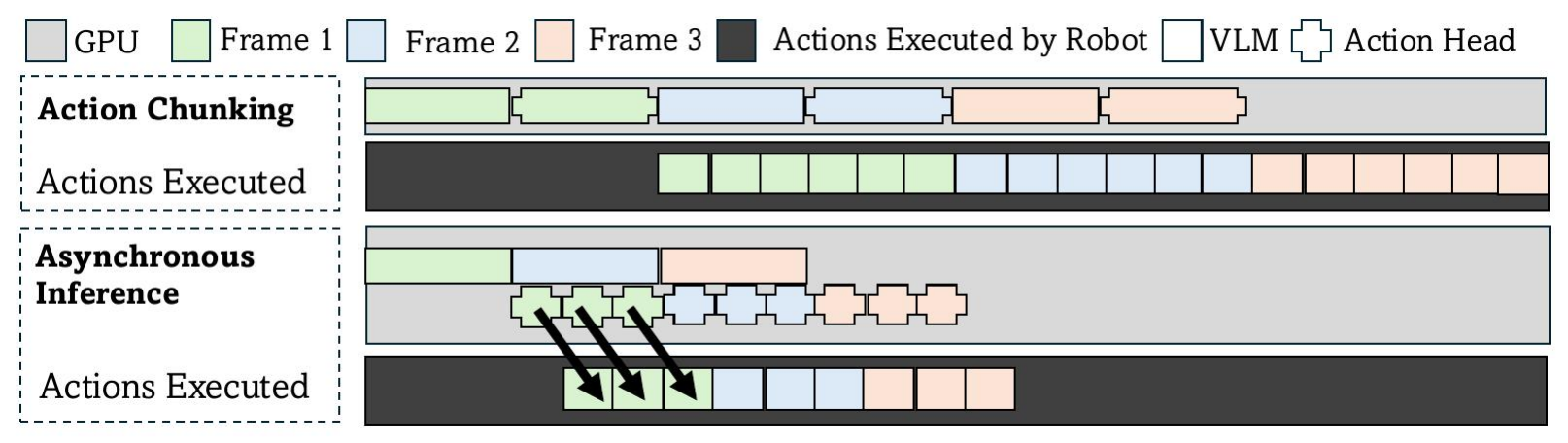}
    \vspace{-0.8cm}
    \caption{Execution timelines for action chunking and asynchronous inference.}
    \vspace{-0.4cm}
    \label{fig:vlaconc}
\end{figure}

To accelerate VLA inference for real-time robotic control~\cite{vlacache}, two complementary techniques are commonly adopted. First, action chunking~\cite{cogact, pi, octo, graspvla} emits multiple future actions, known as an \emph{action chunk}, per inference call, amortizing inference cost across many control steps~\cite{rtc, faster, vlash, a2c2, smolvla}. As illustrated in Figure~\ref{fig:vlaconc}, action chunking allows action execution to overlap with VLA inference. Second, asynchronous inference decouples the update frequencies of the VLM and the action head, allowing the action head to generate multiple action chunks from a single VLM forward pass~\cite{duocore, fis}, reusing the latest available VLM output. As illustrated in Figure~\ref{fig:vlaconc}, multiple action head inference passes run after one VLM inference.
While asynchronous VLAs explicitly parallelize the VLM and action head~\cite{duocore, fis}, in VLAs that employ action chunking, it is also possible to decouple the VLM and action phase~\cite{rtc}. The VLM forward pass for chunk $N{+}1$ can be launched while the action head is still computing chunk $N$, pipelining the two phases across consecutive chunks on the same GPU. This is the execution model we target: the VLM phase of chunk $N{+}1$ runs concurrently with the action phase of chunk $N$ via two CUDA streams on the same GPU. Each action head call still consumes the correct VLM output for its frame, preserving sequential semantics. This is timely because generating fresh action chunks at higher rates is critical for reactive control on dynamic tasks~\cite{rtc, faster, vlash, autohorizon}. However, we find that the naïve approach of assigning the two phases to separate CUDA streams~\cite{cudaapi, realtime_speed} falls short of optimal throughput. We identify the sources of inefficiency in Section~\ref{sec:perfvla}.

\subsection{GPU Architecture}
\label{sec:gpu}
\begin{figure}[htbp]
    \centering
    \includegraphics[width=0.85\linewidth]{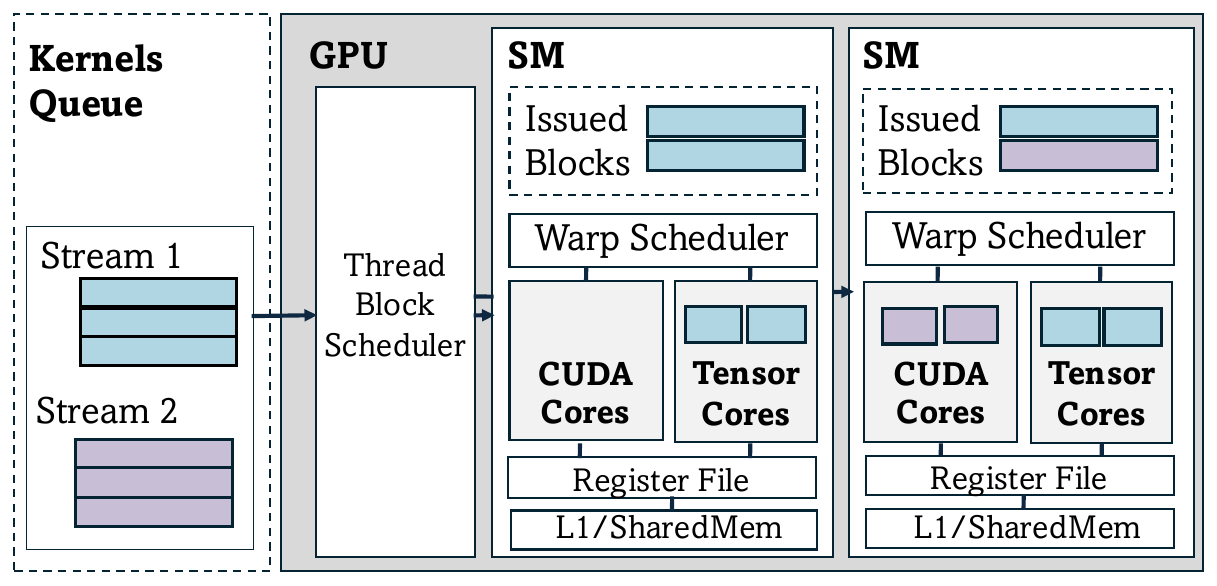}
    \vspace{-0.3cm}
    \caption{High-level overview of Streaming Multiprocessor (SM) and processing cores on modern GPUs.}
    \label{fig:sm-arch}
\end{figure}

Modern GPUs are parallel processors built on a Single Instruction Multiple Threads (SIMT) architecture (See Figure~\ref{fig:sm-arch}). Computation is distributed across an array of Streaming Multiprocessors (SMs). 

\vspace{-0.8em}
\vspace{-1em}

Each SM contains Tensor Cores for dense matrix operations, general-purpose CUDA Cores for scalar arithmetic, a large register file, and a configurable shared memory/L1 cache~\cite{nvidia_ada_whitepaper}.
When a kernel is launched, its threads are grouped into thread blocks, known as Cooperative Thread Arrays (CTAs). The CTA is the fundamental scheduling granularity assigned to an SM, and each resident CTA on an SM is subdivided into warps of 32 threads that execute in lockstep. 
CTAs are dispatched to SMs via a hardware thread block scheduler. However, when scheduling concurrent workloads via CUDA streams~\cite{cudaapi}, this scheduler has two idiosyncrasies~\cite{gpupool, paella}:

\noindent\textbf{(1) Greedy CTA Admission.} The CTA scheduler launches the maximum number of thread blocks per SM permitted by execution-context limits (registers, shared memory, thread slots), so kernels with large grid sizes saturate the per-SM context storage, preventing thread blocks of concurrent kernels from co-residing on the same SM~\cite{hardwarep}. Intra-SM co-execution of multiple concurrent kernels is thus rare when the running kernel has a large grid size~\cite{gpupool, warpslicer, smk}.

\noindent\textbf{(2) Head-of-Line Blocking.} Even when SMs have spare resources along some dimensions (i.e., shared memory, register file, unused thread/block slots), the CTA scheduler does not dispatch thread blocks from another kernel until the first running kernel has fully launched all thread blocks defined by its grid size~\cite{gpupool, paella, gilman_sigmetrics}. When a running kernel saturates one per-SM resource, such as registers, the remaining underutilized resources cannot be claimed by a waiting kernel. As a result, kernels with many thread blocks force smaller concurrent kernels to wait, serializing their execution.

\subsection{Performance Characterization of VLAs on Modern GPUs}
\label{sec:perfvla}

We perform a detailed performance study by profiling~\cite{ncu} two representative VLAs on three modern GPU architectures \textemdash~Ada Lovelace (RTX~4090), Hopper (H100), and Blackwell (RTX~5090). We make three key observations.

\begin{figure}[htbp]
    \centering
    \includegraphics[width=1\linewidth]{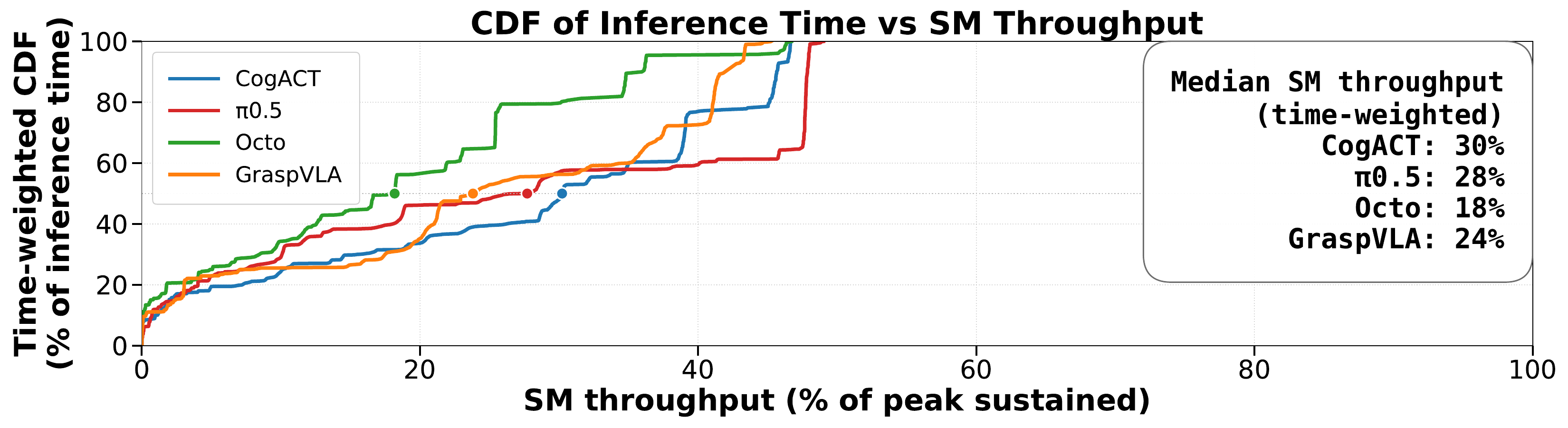}
    \vspace{-0.6cm}
    \caption{CDF of inference time vs. SM throughput for a single-image inference on the RTX~4090.}
    \vspace{-0.2cm}
    \label{fig:smcdf}
\end{figure}

\begin{figure}[htbp]
    \centering
    \includegraphics[width=\linewidth]{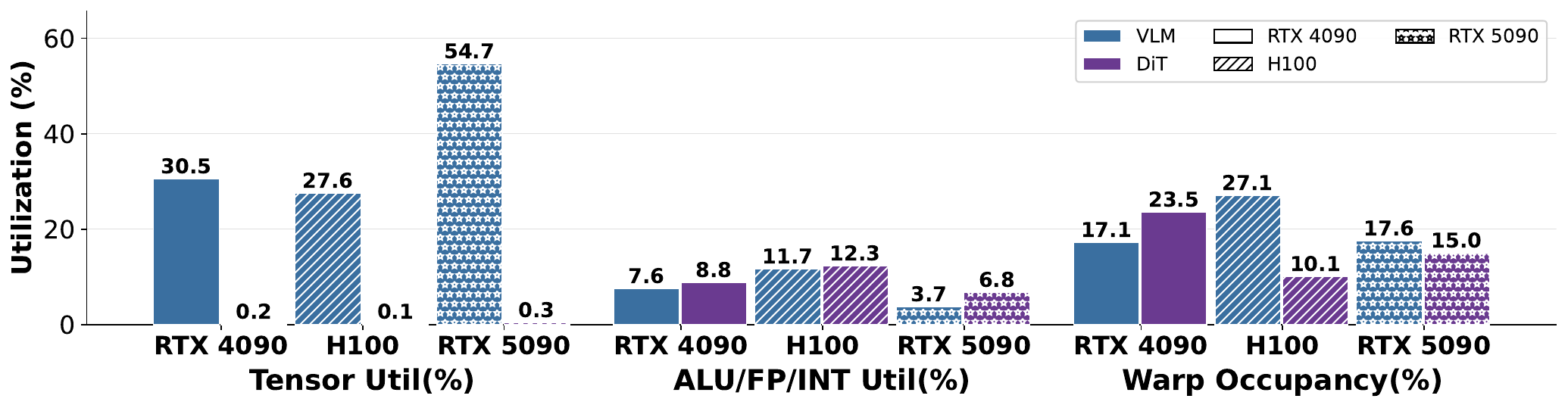}
    \includegraphics[width=\linewidth]{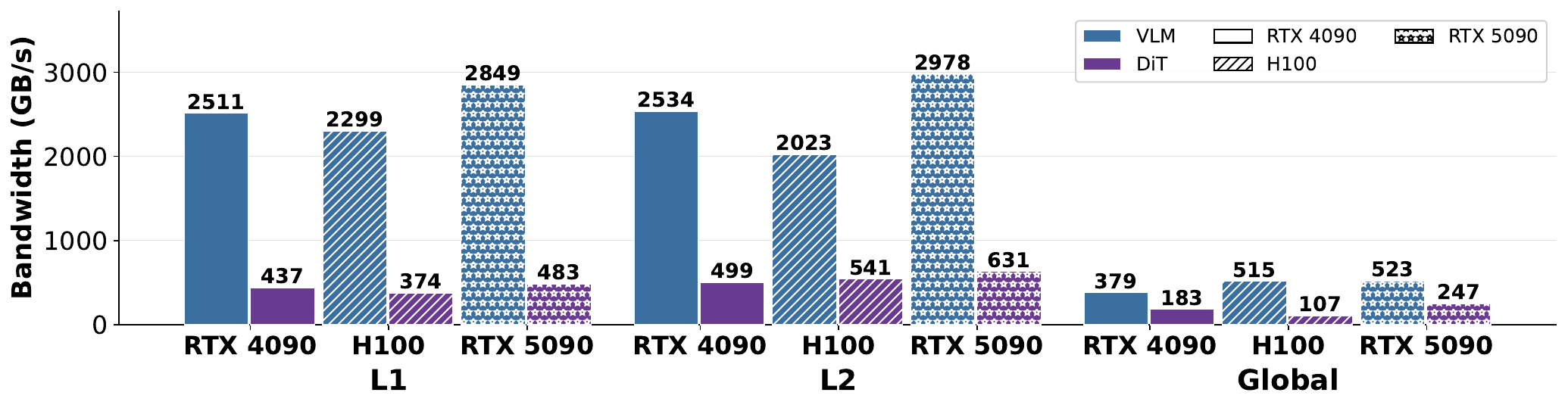}
    \includegraphics[width=\linewidth]{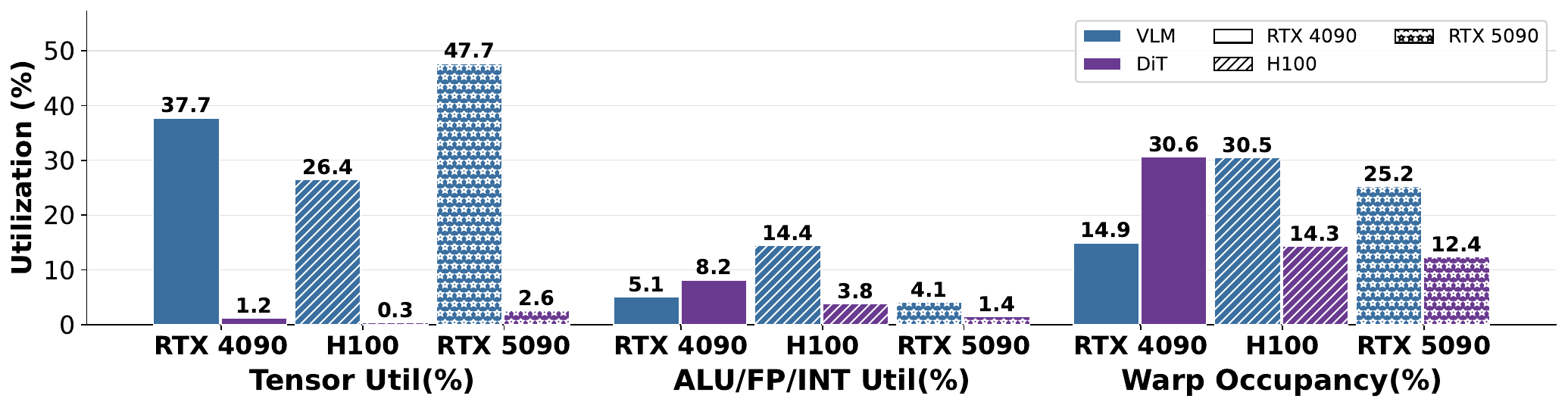}
    \includegraphics[width=\linewidth]{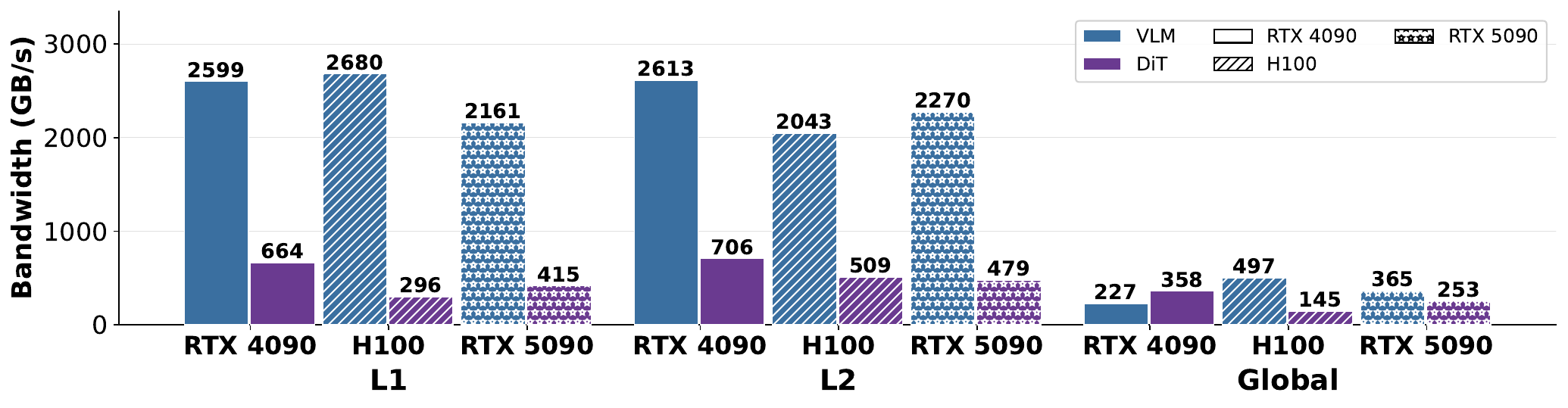}
    \vspace{-0.8cm}
    \caption{Resource utilization profiles of the VLM and Action Head phases for CogACT (top) and $\pi0.5$ (bottom) across RTX 4090, H100, and RTX 5090. 
    }
    \label{fig:tc_ncu_util}
    \vspace{-0.5cm}
\end{figure}

First, VLAs significantly underutilize the GPU during inference (Figure~\ref{fig:smcdf}). 
The median time-weighted SM throughput across all four models ranges from $18\%$ to 30\% of peak sustained throughput. Even the 20–30\% of inference time during which SM utilization is highest never exceeds roughly 50\% of peak. Figure~\ref{fig:tc_ncu_util} illustrates the utilization of Tensor Cores, ALU/FP/INT units, warp occupancy, and memory bandwidth of the VLM and action phases on RTX~4090, RTX~5090 and H100. 
The action phase significantly underutilizes the Tensor Cores, averaging below 3\% utilization. This is because the action phase consists primarily of small matrix operations and element-wise computations that do not effectively leverage Tensor Cores, unlike the VLM phase which has a higher utilization of Tensor Cores because of its large reduced-precision GEMMs. L1 cache, L2 cache, and DRAM bandwidth utilization also differ substantially between the two phases. 

\begin{figure}[tbp]
    \centering
    \includegraphics[width=1\linewidth]{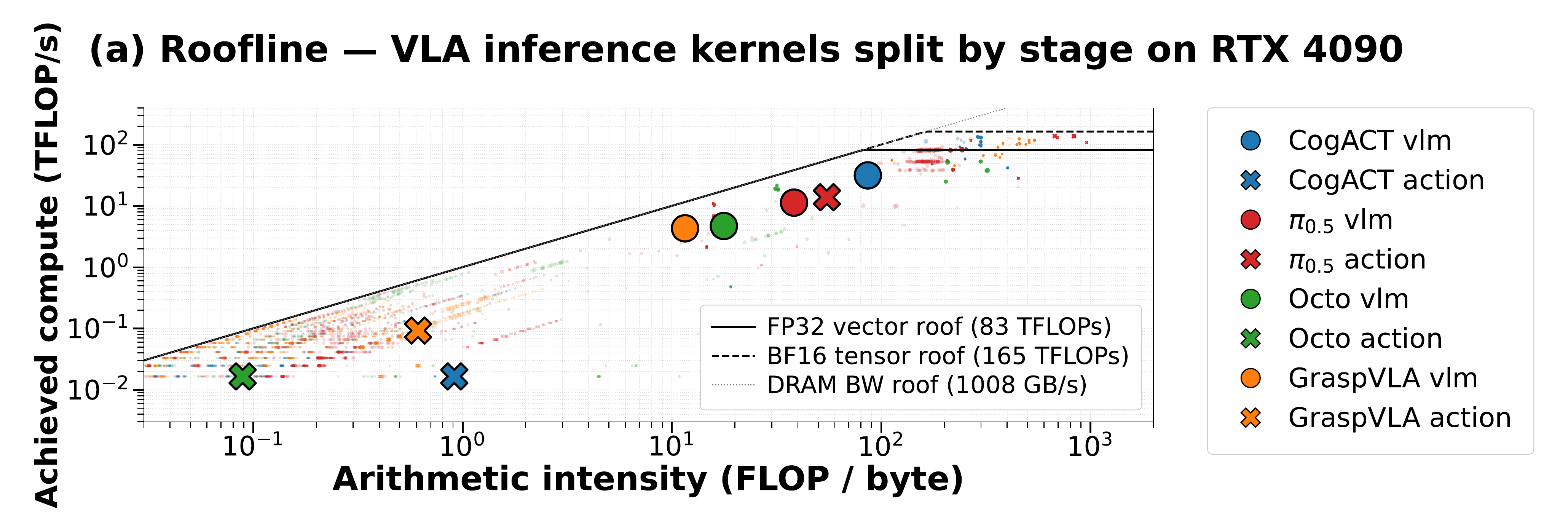}
    \vspace{-0.8cm}
    \caption{Roofline of VLA kernels on RTX 4090. Large markers are time-weighted centroids.}
    \label{fig:roofline}
\end{figure}

Second, the VLM phase  exhibits higher arithmetic intensity than the action phase (Figure~\ref{fig:roofline}). The GEMM kernels of the VLM phase consistently achieve high tensor core utilization in all GPU architectures (37.4\% on average).

Third, when the two phases execute concurrently, the VLM phase runs at approximately the same rate as in isolation; the head-of-line blocking penalty falls entirely on the action phase, as observed in Nsight Systems profile for the four models (Figure~\ref{fig:streamexec}b). 
Given these observations, overlapping the two phases will become beneficial when the penalty on the action head is mitigated.

\subsection{Challenges}
\label{sec:challenges}

\begin{figure}[htbp]
    \centering
    \includegraphics[width=1\linewidth]{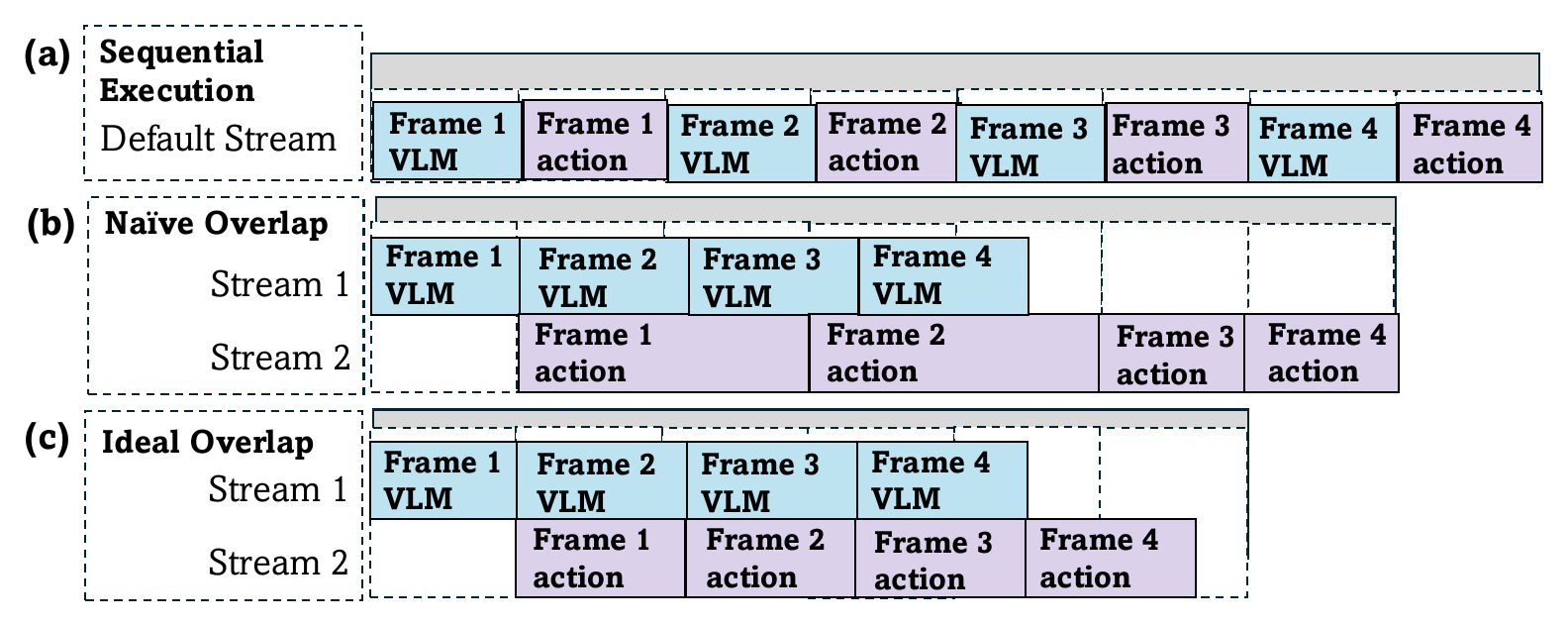}
    \vspace{-0.8cm}
    \caption{Execution patterns for asynchronous VLA inference. (a) Sequential execution serializes the two phases. (b) Naive two-stream overlap leaves the action stream lagging behind. (c) Ideal overlap fully hides action-head latency behind the next frame's VLM.}
    \label{fig:streamexec}
\end{figure}

While both the VLM and action phases underutilize the GPU and stress different functional units, running them concurrently yields negligible throughput gain, as illustrated by the Naive Overlap in Figure~\ref{fig:streamexec}b. This is caused by the head-of-line blocking behavior in GPUs, discussed in Section~\ref{sec:gpu}. Small action-phase kernels are serialized behind the large VLM-phase kernels by the GPU scheduler, even though the functional units needed by the action phase have substantial headroom on every SM throughout the execution~\cite{realtime_speed}.

\begin{figure}[t]
    \centering
    \includegraphics[width=\linewidth]{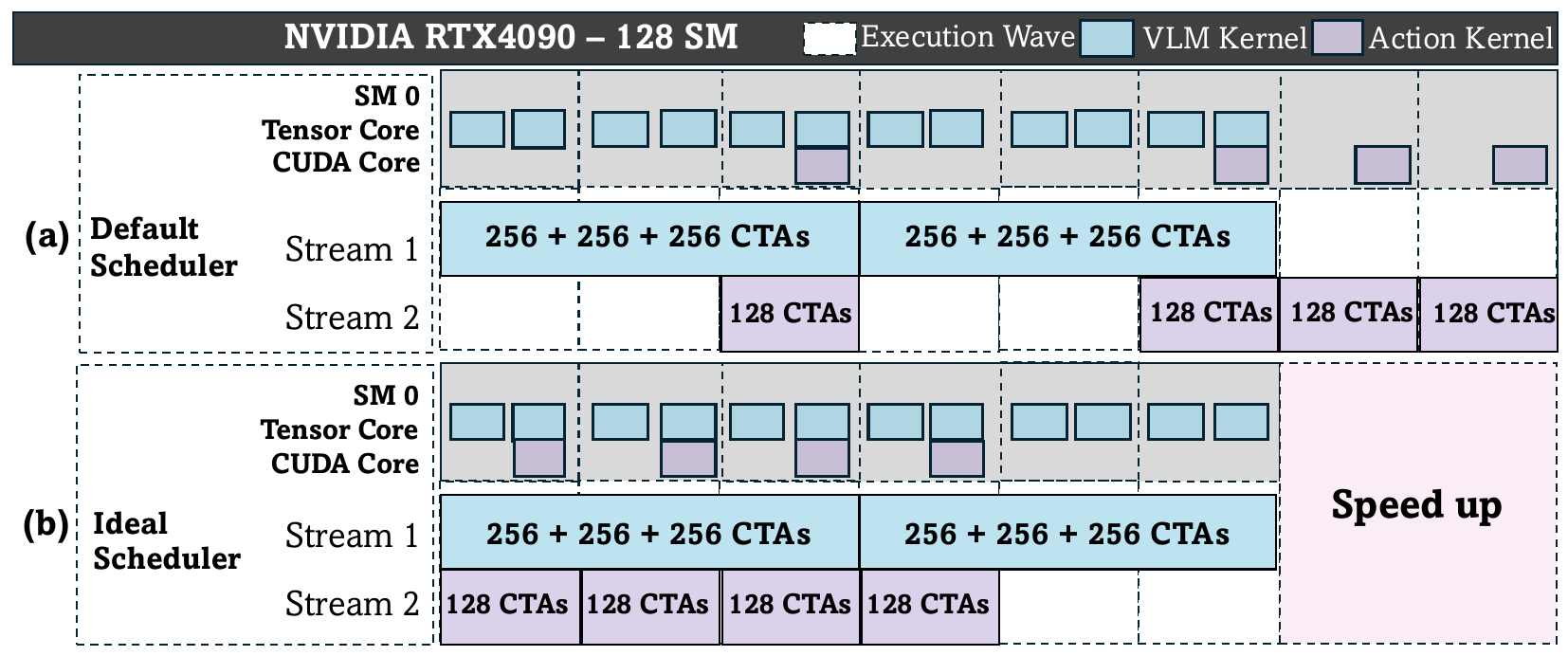}
    \vspace{-0.7cm}
    \caption{
    \textbf{(a) Default GPU scheduling.} Head-of-line blocking behavior. 
    \textbf{(b) Ideal co-scheduling.} Fully overlapped execution within an SM.
    }
    \vspace{-0.6cm}
    \label{fig:timeline}
\end{figure}

This inefficiency arises from the CTA dispatcher's greedy admission policy, illustrated in Figure~\ref{fig:timeline}a. The VLM's GEMM kernels (blue) launch with large grid sizes, i.e., 768 CTAs. Because each SM can host only two such CTAs, given their register footprint, the kernel takes multiple waves (three, in this case) to drain. Meanwhile, the action phase launches a kernel with 128 thread blocks. The thread block dispatcher saturates every SM with two VLM CTAs in the first two waves of execution and admits action phase thread blocks (purple) only in the third wave, due to the head-of-line blocking behavior. Therefore, even when the VLM and action phases are launched in parallel, SMs spend most of the time executing only VLM kernels, saturating the Tensor Cores while leaving the CUDA Cores idle. The action phase does not execute despite ample headroom on the functional units it needs. 

As established in Section~\ref{sec:vla}, modern asynchronous VLA inference~\cite{rtc, pi} decouples the two phases across frames: the VLM encodes frame $N{+}1$ while the action head generates actions for frame $N$. 
Figure~\ref{fig:streamexec} illustrates the resulting execution patterns. Under sequential execution (a), the two phases serialize on a single stream. Under naive two-stream overlap (b), the VLM stream runs continuously, but head-of-line blocking at the GPU scheduler keeps the action phase kernels queued behind the VLM kernels, so action chunks are emitted at half the rate they would be otherwise, as observed in the Nsight System profiles. Under ideal overlap (c), each action-head invocation fully overlaps with the next frame's VLM phase, and chunks are emitted at a higher rate. The gap between (b) and (c) is the inefficiency we target, and closing it is challenging for two reasons specific to VLA workloads:
(1) The VLM kernel with which a given action-head kernel overlaps is non-deterministic; there is no guarantee that a specific kernel from the VLM phase will overlap with a kernel from the action head phase since both kernels shift at microsecond granularity. This is because the VLM and action streams are launched independently and progress at different rates, so given a dispatched action-head kernel, the VLM kernel that's resident on the GPU at the same time as that action-head kernel is different across frames (Figure~\ref{fig:streamexec}b). The dispatcher therefore sees a kernel mix that is a constantly shifting combination of one of many VLM kernels paired with one of many action-head kernels.
(2) The action head is composed of micro-kernels. As shown in Table~\ref{tab:kernel-5k-cycle-cudagraph}, the action phase consists of hundreds of small kernels with median duration on the order of a few microseconds. Combined with (1), this means the co-running kernel mix turns over at microsecond timescales, faster than any epoch-based profiling tool can track and far shorter than the millisecond-scale cost of preemption-based partition resizing, which must flush resident thread-block state or halt SMs to redistribute resources~\cite{reef}. Any scheduler whose decision latency exceeds the duration of the kernels it is scheduling will spend more time deciding than computing.

\begin{table}[hb]
\centering
\footnotesize
\caption{Execution profile of micro-kernels across different VLA workloads after applying a CUDA Graph capture.}

\vspace{-0.3cm}
\label{tab:kernel-5k-cycle-cudagraph}
\begin{tabular}{@{}lrrrrr@{}}
\toprule
\textbf{Workload} & \textbf{Median} & \textbf{$<$5K} & \textbf{$<$10K} & \textbf{$<$25K} & \textbf{$<$50K} \\
                  & \textbf{(cyc)}  & \textbf{(\%)}  & \textbf{(\%)}   & \textbf{(\%)}   & \textbf{(\%)}   \\
\midrule
CogACT \cite{cogact}      &  7,156  & 18.9  & 67.2   & 88.9   &  95.2   \\
Octo \cite{octo}          &  6,832  & 44.3  & 69.8   & 85.4   & 100.0   \\
$\pi_{0.5}$ \cite{pi}          &  9,348  & 17.9  & 51.7   & 73.7   &  90.2   \\
GRASP-VLA \cite{graspvla} & 12,407  & 23.9  & 46.3   & 77.6   &  88.4   \\
\bottomrule
\end{tabular}
\end{table}

A VLA-aware co-scheduler must therefore be \emph{micro-kernel responsive}: capable of reacting at microsecond granularity without paying preemption overhead or underutilizing resources due to stale decisions. 
Our goal in this work is to achieve the overlap depicted in Figure~\ref{fig:timeline}b, approaching the ideal pattern of Figure~\ref{fig:streamexec}c.

\subsection{Limitations of Existing Work}

\begin{table}[t]
\centering
\caption{Qualitative comparison of prior GPU co-scheduling frameworks}
\vspace{-0.4cm}
\label{tab:qualitative_comparison}
\renewcommand{\arraystretch}{1.15}
\setlength{\tabcolsep}{4pt}
\footnotesize
\begin{tabular}{@{} l l c c c @{}}
\toprule
\textbf{Scheduler} &
\textbf{Mechanism} &
\textbf{\shortstack{Unit\\Aware}} &
\textbf{\shortstack{Micro-K.\\Resp.}} &
\textbf{\shortstack{Black-\\Box}} \\
\midrule
SMK \cite{smk}                  & HW Ctx Switch   & $\times$     & $\times$     & $\checkmark$ \\
Warped-Slicer \cite{warpslicer} & Epoch + Static  & $\times$     & $\times$     & $\checkmark$ \\
Maestro \cite{Maestro}          & Epoch + 2-Stack & $\times$     & $\times$     & $\checkmark$ \\
GPUPool \cite{gpupool}          & ML + TB/SM Cap  & $\checkmark$ & $\times$     & $\checkmark$ \\
Kitsune \cite{kitsune}          & Dataflow + L2 Q & $\checkmark$ & $\checkmark$ & $\times$     \\
\midrule
\textbf{\pname{} (Ours)}        & \textbf{Dynamic Score} & $\checkmark$ & $\checkmark$ & $\checkmark$ \\
\bottomrule
\end{tabular}

\vspace{2pt}
\footnotesize\textit{Unit Aware} = TC vs.\ ALU execution-unit aware;
\textit{Micro-K.\ Resp.} = micro-kernel responsive;
\textit{Black-Box} = supports unmodified kernels.
\vspace{-0.5cm}
\end{table}

CUDA stream priorities affect only dispatch selection~\cite{gilman_sigmetrics, hardwarep}, not admission, so head-of-line blocking persists. MPS~\cite{mps} and MIG~\cite{mig} target multi-tenant isolation across contexts and processes, not overlapping phases within a single pipeline.

Prior hardware-modification approaches to GPU scheduling, as discussed below, are insufficient for VLA inference. Table~\ref{tab:qualitative_comparison} evaluates the five closest prior frameworks against three essential properties that VLA inference must satisfy: 
\emph{(i) Execution-unit awareness.} The scheduler must distinguish Tensor Core demand from CUDA-core and memory-bandwidth demand at dispatch time and avoid head-of-line blocking. In the VLA setting, the VLM phase is dominated by Tensor Core matrix multiplications while the action phase relies on CUDA-core arithmetic (Section~\ref{sec:perfvla}). Co-locating thread blocks from both phases on the same SM would saturate disjoint execution units, raising aggregate utilization.
(ii) \emph{micro-kernel responsiveness.} The scheduler must react to VLA's rapidly shifting kernel mix at microsecond granularity. Preemption-based techniques cause significant overhead for micro-kernels from flushing resident thread-block state.
(iii) \emph{black-box kernel compatibility.} The scheduler should work with unmodified closed-source vendor binaries such as cuBLAS and cuDNN~\cite{cublas, cudnn}. The Primary Mechanism column captures the design philosophy each system adopts to approach these goals. As Table~\ref{tab:qualitative_comparison} shows, every prior system fails on at least one axis.

\begin{figure*}
    \centering
    \includegraphics[width=1\linewidth]{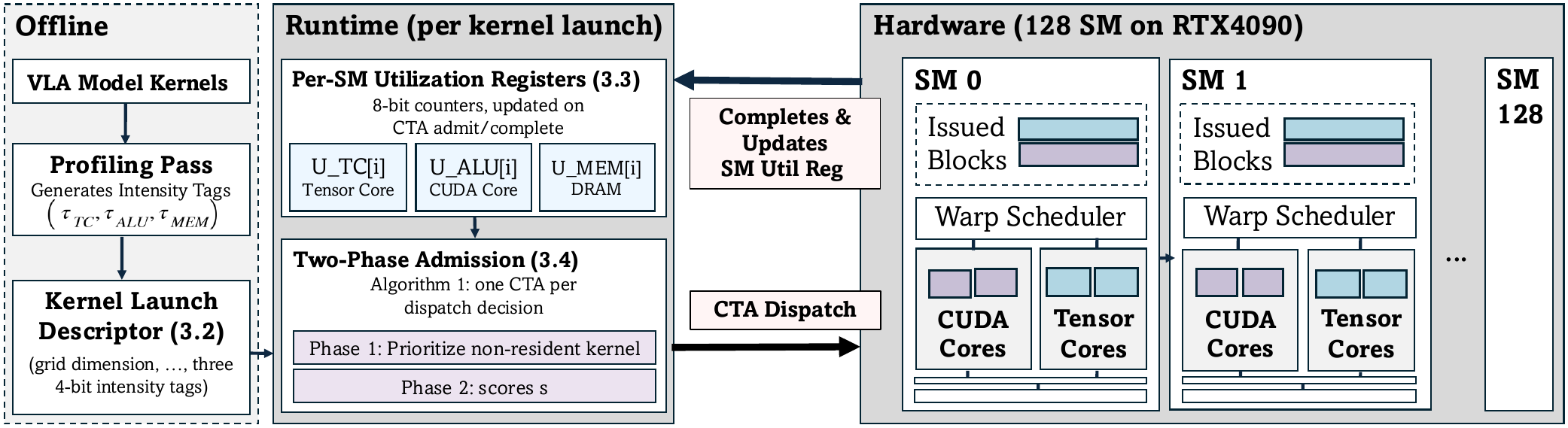}
    \vspace{-10pt}
    \caption{High-level overview of \pname{} design.}
    \vspace{-10pt}
    \label{fig:system-design}
\end{figure*}

\textbf{Simultaneous Multikernel (SMK)}~\cite{smk} co-locates kernels on a single SM by partitioning registers, shared memory, and thread slots, but relies on preemption via Partial Context Switching, which drains in-flight instructions from an SM's pipelines and spills state to device memory. This overhead is prohibitive when a substantial fraction of VLA kernels complete in under 5K cycles (Table~\ref{tab:kernel-5k-cycle-cudagraph}).
\textbf{Maestro}~\cite{Maestro} extends SMK by measuring co-located performance: it tests candidate partitions on designated Trial SMs during one epoch and applies the best-observed ratio to Follower SMs in the next. However, similar to SMK, Maestro resizes partitions by preempting SMs, introducing the same overhead that is incompatible with VLA's microsecond-scale kernel changes. 
\textbf{GPUPool}~\cite{gpupool} modifies the thread block scheduler to bypass head-of-line blocking, but enforces a single static TBs/SM cap per application. This creates a static cap that cannot capture the diverse resource demands of rapidly changing VLA kernels, and it blocks the scheduler from filling idle units during transient VLM stalls.
\textbf{Kitsune}~\cite{kitsune} is execution-unit aware and preemption-free, but requires rewriting kernels with atomic L2 synchronization primitives—incompatible with closed-source operators such as cuBLAS and cuDNN that dominate the VLM phase.
\textbf{Warped-Slicer}~\cite{warpslicer} uses online profiling to determine the number of thread blocks each co-located kernel may occupy on a shared SM. However, its design is mismatched to VLA workloads. First, non-deterministic launch times produce shifting VLM/action kernel pairings, forcing repeated invocation of its online profiling path; avoiding this would require either exhaustive offline cross-profiling of all kernel pairs or an online profile cache, neither of which the original design details. Second, VLA workloads are dominated by sub-5K-cycle micro-kernels (Table~\ref{tab:kernel-5k-cycle-cudagraph}), which are shorter than Warped-Slicer's 5K-cycle profiling window and therefore fall back to its default partitioning. Since Warped-Slicer is non-preemptive, we include it as a baseline in Section~\ref{sec:eval}.

Software-based approaches fail similarly. Orion's~\cite{orion} runtime approach consumes offline profiles at host-side whole-kernel submission, so it cannot react to the dynamic resource utilization change of each SM or guarantee that CTAs from different kernels co-reside on the same SM. In addition to not being applicable to closed-source kernels, Tacker~\cite{tacker} and Adaptive Kernel Fusion~\cite{aker} statically divide each fused thread block between the two kernels. If their execution times differ, resources assigned to the shorter kernel may become idle, and its dependent successor cannot begin until the entire fused kernel completes. 

To conclude, VLAs demonstrate three characteristics that motivate a new design: 
\emph{(i)}~execution-unit awareness,
\emph{(ii)}~microsecond-scale kernel responsiveness,
and \emph{(iii)}~black-box vendor reliance.
This motivates a lightweight CTA dispatcher that accepts per-kernel intensity tags, maintains per-SM utilization counters, and scores pending CTAs for complementarity without kernel modification or warp-scheduler changes.

\section{Our Approach: \pname{}}
\label{sec:design}

\subsection{Overview}

We propose \pname{}, a GPU scheduling framework that enables high-throughput VLA inference. An overview of \pname{} is shown in Figure~\ref{fig:system-design}. \pname{} has two phases. 
The \emph{offline profiling phase} profiles each kernel's per-CTA demand on the tensor, ALU, and memory pipelines and encodes it as an \textit{intensity descriptor} (Section~\ref{sec:kerprofile}). The intensity descriptor is computed once per kernel, stored alongside its binary, and read by the hardware scheduler at launch with the kernel's launch descriptor.
The \emph{online scheduling phase} includes two components: 
(1) per-SM utilization registers (Table~\ref{tab:utilization-registers}) that track aggregate execution-unit load, are updated on CTA dispatch and completion (Section~\ref{sec:smutil}); and 
(2) a dispatching algorithm that selects which pending kernel's thread block to issue by jointly considering kernel intensity descriptors and per-SM utilization registers, co-locating thread blocks with complementary resource demands on the same SM (Section~\ref{sec:scoring}). 

\pname{} addresses both challenges identified in Section~\ref{sec:challenges}:
\textbf{Addressing Challenge 1: Head-of-line blocking.} \pname{} resolves head-of-line blocking with two mechanisms in its dispatch policy (Algorithm~\ref{alg:admission}). Phase~1 removes head-of-line blocking by giving admission priority to kernels that have no CTA resident on the target SM, so a waiting kernel is admitted as soon as the SM has feasible capacity rather than after the running kernel has drained its grid. This guarantees concurrent execution on an SM among kernels. 
Phase~2 reads the intensity descriptor of each kernel (Section~\ref{sec:kerprofile}) and per-SM utilization registers (Section~\ref{sec:smutil}) to fill remaining SM capacity with thread blocks that best complement the pipelines currently underused on that SM. Execution-unit awareness thus emerges as a property of removing head-of-line blocking for VLAs.
\textbf{Addressing Challenge 2: Responsiveness to microsecond-scale kernels.} \pname{}'s dispatch policy, Algorithm~\ref{alg:admission}, reads the kernel intensity descriptors and per-SM utilization registers that are already available to the dispatcher at runtime. One invocation of Algorithm~\ref{alg:admission} evaluates K candidate kernels once and commits one CTA; it contains no iteration over CTA slots and no additional arbitration round. The cost of this decision is therefore independent of kernel duration and  SM occupancy, making it applicable to kernels of any length.

\subsection{Kernel Resource Profile Descriptor}
\label{sec:kerprofile}

\subsubsection{Kernel Launch Descriptor Extension}
Prior empirical studies and our profiling of GPUs identify that thread-block placement policies are based only on static occupancy constraints (e.g., register and shared-memory limits), ignoring which functional units (e.g., Tensor Cores, FP32 ALUs) each kernel stresses. This gap spans both sides of the hardware-software interface: the scheduler cannot observe per-kernel functional-unit usage, and existing APIs offer no channel to communicate it.

To bridge this gap, we extend the kernel launch descriptor, which is the metadata block passed from the host to the GPU at kernel launch time. The kernel launch descriptor already includes grid dimensions, per-thread register count, and shared memory allocation. We add three 4-bit fields to it that encode per-kernel execution-unit intensity values $(\tau_{TC}, \tau_{ALU}, \tau_{MEM})$ as defined in Table~\ref{tab:intensity-fields}. 
Each $\tau_i$ is the fraction of the corresponding peak throughput consumed per CTA: $\tau_{TC}$ for Tensor Cores, $\tau_{ALU}$ for CUDA Cores, and $\tau_{MEM}$ for DRAM bandwidth. Together, the three fields form a 12-bit signature that the thread-block scheduler reads at dispatch time to determine whether thread blocks from different kernels will share an SM efficiently or contend for the same units.
This 12-bit annotation can be communicated via an extension to existing per-kernel attribute APIs~\cite{cudaapi}. The change is confined to the driver.

\begin{table}[htbp]
\centering
\footnotesize
\caption{A kernel's Intensity Descriptor. Each field (4 bits) encodes a normalized intensity level in the range 0--15.}
\vspace{-0.3cm}
\label{tab:intensity-fields}
\begin{tabular}{@{}lp{0.88\linewidth}@{}}
\toprule
\textbf{Field} & \textbf{Description} \\
\midrule
$\tau_{TC}$ & Tensor Core (TC) pipeline intensity. Fraction of peak TC throughput that CTAs consume. \\
$\tau_{ALU}$ &  ALU pipeline intensity. Fraction of peak ALU (CUDA Core) throughput consumed per CTA. \\
$\tau_{MEM}$ &  Memory bandwidth intensity. Fraction of peak DRAM bandwidth consumed per CTA. \\
\bottomrule
\end{tabular}
\vspace{-0.3cm}
\end{table}

\subsubsection{Profile Metadata}
The kernel intensity descriptors are obtained through a one-time offline profiling pass, performed once per model architecture at deployment time. Using hardware performance counters~\cite{ncu}, each unique kernel is executed in isolation and profiled. For each distinct kernel, we record the achieved (1) tensor-core FLOP/s (FP16/BF16 matrix-multiply) as a fraction of the GPU's peak tensor-core FLOP/s; (2) FP32 FLOP/s on the CUDA Cores as a fraction of peak CUDA-core FP32 FLOP/s; and (3) DRAM throughput as a fraction of peak sustained bandwidth. The utilization numbers are then quantized to the 0-15 integer range to produce the corresponding $\tau$ fields in Table~\ref{tab:intensity-fields}. The triplet $(\tau_{TC}, \tau_{ALU}, \tau_{MEM})$ is stored in a file and injected at kernel launch by a wrapper around the runtime kernel launch API.

For the VLA models evaluated in this paper, profiling all unique kernels typically takes two to three hours, but can take up to 6 hours depending on model size. This profiling cost happens \emph{only once} per model, which is negligible in our target scenario: a VLA model is configured at deployment time and then serves millions of inference requests over the robot's operational lifetime, amortizing the one-time cost across the entire deployment. 

Moreover, VLAs' input shapes are mainly fixed as their batch size, camera count, image resolution, action dimensions, data-types are determined at deployment and do not vary across inference calls. Input text-length is the only dimension that varies. We therefore profile a small set of text-length buckets to account for closed-source libraries selecting different kernels across input sequence lengths. If an unseen configuration is nevertheless dispatched at runtime and \pname{} finds no matching descriptor, it will simply fall back to the baseline admission policy for that kernel. 

\subsection{Per-SM Utilization Registers}
\label{sec:smutil}
\subsubsection{State Registers} 
We introduce three per-SM utilization registers, $U_{TC}[j]$, $U_{ALU}[j]$, and $U_{MEM}[j]$ (Table~\ref{tab:utilization-registers}), one per execution-unit class (TC, ALU, MEM). Each register holds the sum of the corresponding $\tau$ field over all thread blocks currently resident on SM $j$. That is, $U_{TC}[j] = \sum_{b} \tau_{TC}(b)$ across resident blocks $b$ on SM $j$, thereby giving an estimate of how heavily the Tensor Core execution unit is loaded. The three per-SM utilization registers are updated incrementally: a block's $\tau$ values are added on dispatch and subtracted on completion. The scheduler reads the triplet $(U_{TC}[j], U_{ALU}[j], U_{MEM}[j])$ of SM $j$ at dispatch time to decide which kernel's thread block will be able to run on SM $j$ without contending for units that are already saturated.
\begin{table}[htbp]
\vspace{-0.2cm}
\centering
\footnotesize
\caption{Per-SM 8-bit utilization registers (range $[0, 255]$) maintained by the
scheduler.}
\vspace{-0.3cm}
\label{tab:utilization-registers}
\begin{tabular}{@{}lp{0.82\linewidth}@{}}
\toprule
\textbf{Register} & \textbf{Description} \\
\midrule
$U_{TC}[j]$  & Estimated Tensor Core utilization on SM $j$. \\
$U_{ALU}[j]$ & Estimated ALU pipeline utilization on SM $j$. \\
$U_{MEM}[j]$ & Estimated memory bandwidth utilization on SM $j$. \\
\bottomrule
\end{tabular}
\vspace{-0.3cm}
\end{table}

\subsubsection{Register Update Mechanism}

The thread block scheduler already maintains per-SM resource accounting to determine admission feasibility: each SM's use of registers, shared memory, and thread block slots is tracked so that, at dispatch time, the scheduler checks whether a thread block of kernel $K$ fits on $SM_j$, and upon thread block completion, the counters are updated to account for the released resources~\cite{gilman_sigmetrics}. 
We extend this resource tracking mechanism present in the hardware scheduler. A thread block's $\tau$ values are added to $U_{TC}[j]$, $U_{ALU}[j]$, and $U_{MEM}[j]$ when the thread block is dispatched to SM $j$, and subtracted when the thread block completes. In doing so, we reuse the same thread block completion signals that already trigger the release of registers, shared memory, and thread block slots back to the SM's free pool. Counters saturate at 255: an addition that would overflow is capped at 255, signaling the scheduler that the SM is fully loaded for that hardware resource. Since 0-255 is a large range, even if resources are capped, it would minimally affect the effectiveness of the scheduler. In addition, saturation is rare in practice, since it requires 17 co-resident thread blocks each at peak tensor-core intensity. This situation does not occur in the workloads we evaluate, where kernels at $\tau_{TC}$ = 15 have register footprints that admit far fewer than 17 CTAs per SM.

\begin{algorithm}[htb]
\scriptsize
\caption{Single-CTA Admission Decision on SM $j$}
\label{alg:admission}
\begin{algorithmic}[1]
\Statex \textbf{State (persistent, per SM $j$):}
\Statex \quad $\text{CTA\_count}(k,j)$: CTAs of kernel $k$ resident on SM $j$
\Statex \quad $U_{TC}[j], U_{ALU}[j], U_{MEM}[j]$: estimated utilization
\Statex
\Statex \textit{Invoked once per CTA-admission decision; admits at most one CTA.}
\Statex \textbf{Phase 1: Prioritize one unrepresented feasible kernel}
\For{each kernel $k$ with pending CTAs, in dispatcher order}
    \If{$\text{CTA\_count}(k,j) = 0$}
        \Comment{$k$ is not resident on SM $j$}
        \If{$\textsc{Feasible}(k,j)$}
            \Comment{occupancy constraints permit admission}
            \State $\textsc{Admit}(k,j)$
            \State Update resident-CTA count and per-SM utilization registers
            \State \textbf{return}
            \Comment{commit exactly one admission}
        \EndIf
    \EndIf
\EndFor
\Statex
\Statex \textbf{Phase 2: Select one feasible kernel by complementarity}
\Statex \textit{Executed only when Phase 1 admits no CTA.}
\State $\text{best\_score} \gets -1$
\State $\text{best\_kernel} \gets \textsc{Null}$
\For{each kernel $k$ with pending CTAs, in dispatcher order}
    \If{$\lnot\;\textsc{Feasible}(k,j)$}
        \State \textbf{continue}
        \Comment{insufficient occupancy resources}
    \EndIf
    \State $s \gets (U_{\max} - U_{TC}[j]) \cdot \tau_{TC}(k)$
    \Statex \hspace{3.2em}
        $+ \;(U_{\max} - U_{ALU}[j]) \cdot \tau_{ALU}(k)$
    \Statex \hspace{3.2em}
        $+ \;(U_{\max} - U_{MEM}[j]) \cdot \tau_{MEM}(k)$
    \If{$s > \text{best\_score}$}
        \Comment{strict comparison favors first-seen kernel}
        \State $\text{best\_score} \gets s$
        \State $\text{best\_kernel} \gets k$
    \EndIf
\EndFor
\If{$\text{best\_kernel} = \textsc{Null}$}
    \Comment{no pending kernel is feasible}
    \State \textbf{return}
\EndIf
\State $\textsc{Admit}(\text{best\_kernel},j)$
\State Update resident-CTA count and per-SM utilization registers
\State \textbf{return}
\Comment{commit exactly one admission}
\Statex
\Statex \textbf{On CTA completion} $(k,j)$\textbf{:}
\Statex \quad $\text{CTA\_count}(k,j) \gets \text{CTA\_count}(k,j) - 1$
\Statex \quad Subtract $\tau_{TC}(k), \tau_{ALU}(k), \tau_{MEM}(k)$ from $U_{TC}[j], U_{ALU}[j], U_{MEM}[j]$
\end{algorithmic}
\end{algorithm}

\subsection{Execution-Unit-Aware CTA Dispatch}
\label{sec:scoring}
\subsubsection{Scoring Function}
Co-locating kernels on an SM is beneficial only if their thread blocks need to use different hardware resources. Otherwise, contention may degrade performance. The thread block dispatcher must therefore (i)~detect when a pending kernel's utilization of hardware resources complements an SM's current hardware resource utilization, and (ii)~keep kernels co-resident on the same SM rather than draining one before admitting the next, in order to eliminate head-of-line blocking.
For example, when a tensor-core-intensive kernel and a memory-intensive kernel are pending at the same SM, the dispatcher should admit thread blocks from each kernel in proportion to the SM's headroom along the corresponding pipelines, producing a co-resident mix that balances utilization across execution units. \pname{} enables such kernel dispatching via Algorithm~\ref{alg:admission}, applied independently to each SM. 
Algorithm 1 is invoked once for each ordinary CTA-admission decision on SM $j$ and commits at most one CTA. It does not fill all available CTA slots within one invocation. If the SM has sufficient capacity after an admission, the work distributor invokes Algorithm 1 again during a subsequent admission cycle, using the updated state registers. 

\textbf{Phase 1: co-residency priority (Lines~1--8).} Phase 1 ensures a kernel whose intensity profile scoring marginally but consistently below a co-running kernel's is not starved. Phase 2 could repeatedly select a same kernel and prevent kernel co-residency. Phase~1 avoids this by unconditionally admitting one thread block from each active kernel, guaranteeing forward progress on every SM. It collects the kernels that are pending, feasible, and have no CTA on SM~$j$, then picks one by round-robin. The count $C[k,j]$ persists across invocations, so the next call sees that kernel as resident and picks a different one. Thus, every kernel that fits on SM~$j$ makes progress.

\textbf{Phase 2: complementarity-driven selection (Lines~10--27).} Phase~2 runs only when Phase~1 has not admitted any kernel: either every feasible kernel already has a CTA on SM~$j$, or the kernel does not fit on SM $j$ due to resource constraints. The dispatcher scans the pending kernels and skips any that do not fit in SM~$j$'s registers or shared memory (Line~13). For each remaining kernel, Line~16 computes a \emph{compatibility score} $s$ as the dot product of the kernel's descriptor with the SM~$j$'s headroom, stored in the per-SM state registers.
Note $U_{\max}$ is the per-pipeline saturation cap, so $U_{\max} - U_{*}[j]$ measures the headroom remaining on pipeline $*$. $s$ is large when the kernel demands the pipelines SM~$j$ is currently underusing. Lines~17--20 track the
highest-scoring kernel, with the strict inequality breaking ties. The winning thread block is admitted (Line~25), SM~$j$'s resident-CTA count and utilization registers are updated (Line~26), and the
algorithm returns. Re-scoring happens at the next admission decision on SM~$j$, against the updated headroom.

\vspace{-0.7em}
\subsubsection{Implementation Overhead}
\pname{} requires modest, localized changes to the GPU stack. 
\textbf{Compiler and runtime.} The kernel intensity descriptor is a 12-bit field appended to the kernel binary's metadata, read alongside the existing launch descriptor. 
\textbf{Hardware scheduler.} \pname{} extends the thread-block dispatcher in two ways: 
\textit{(i) Per-SM scheduling state:} three 8-bit utilization registers per SM (24 bits) join the dispatcher's per-SM status registers, updated by 8-bit saturating adders on CTA dispatch/completion using existing signals. Algorithm~1 additionally requires the resident-CTA count $\text{CTA\_count}(k,j)$; with $K=2$ and at most 32 resident CTAs per SM (Table~5), this is two 6-bit counters, for 36 bits of state per SM. On an RTX 5090 (170 SMs), this is 765 bytes, which is negligible. The same counters are read directly by the fairness-augmented variants of Section~4. 
\textit{(ii) Two-phase admission logic:} one invocation of Algorithm~1 iterates once over the $K$ active kernels, where $K$ is the number of concurrent streams, and commits exactly one \textsc{Admit}. The per-kernel score $s$ is a three-term dot product of 4-bit intensity fields against 8-bit headroom values, computed in a small combinational block (six $8{\times}4$-bit multipliers for $K=2$, two-level adder trees, ${\sim}1500$ gates, ${<}0.01\%$ of the arbitration logic area) running in parallel with the baseline register file and shared-memory feasibility check. The logic scales with $K$, not with the 24--32 CTA slots per SM; filling multiple slots requires multiple arbitration cycles. No extra memory access, pipeline stage, or arbitration round is added, so dispatch completes within single-digit-cycle budget. 
\section{Evaluation Methodology}

\noindent\textbf{Simulation Methodology.}
We implement \pname{} using Accel-Sim~\cite{accelsim}, a GPU simulator. We use two configurations that simulate NVIDIA RTX~4090 and RTX~5090 architectures. Table~\ref{tab:method} lists the microarchitectural parameters of each configuration~\cite{accelsim, nvblackwell}.

\vspace{-0.4em}
\begin{table}[htbp]
\footnotesize
    \caption{GPU Microarchitectural Parameters in Accel-Sim}
    \vspace{-8pt}
    \begin{tabular}{lll}
        \toprule
        \textbf{Parameter} & \textbf{NVIDIA RTX 4090} & \textbf{NVIDIA RTX 5090} \\
        \midrule
        Generation & Ada Lovelace & Blackwell \\
        SM Clock Frequency & 2520 MHz & 2407 MHz \\
        SMs/GPU & 128 & 170 \\
        Warp Schedulers/SM & 4 & 4 \\
        Warp Scheduling Policy & GTO & GTO \\
        Max \#Thread Blocks/SM  & 24            & 32 \\
        Max Shared Memory/SM    & 100 KB        & 128 KB \\
        Max Registers/SM & 65536 & 65536 \\
        Register File Size & 256 KB & 256 KB \\
        L1 Cache Size & 128 KB & 128 KB \\
        \bottomrule
    \end{tabular}
    \label{tab:method}
\end{table}
\vspace{-0.4cm}

\noindent\textbf{Workloads.}
We evaluate \pname{} using four VLA models that span a range of architectures and action-head designs: \textbf{CogACT}~\cite{cogact}, \textbf{Octo}~\cite{octo}, \textbf{$\pi_{0.5}$}~\cite{pi}, and \textbf{GRASP-VLA}~\cite{graspvla}. We trace the GPU kernels of each model on physical RTX~4090 and RTX~5090 GPUs using NVBit~\cite{nvbit}, then replay the traces in Accel-Sim~\cite{accelsim}. We perform inference on images from the Open X-Embodiment Dataset~\cite{openx}.

\noindent\textbf{\pname{} Implementations.}
We evaluate the baseline \pname{} design (Section~\ref{sec:design}) \textemdash~\pname{}, and two variants that modify the Phase-2 scoring function in Algorithm~\ref{alg:admission}: (1) \pnameg{} scales $s$ by a grid-progress factor, $s_{GRID} = s \cdot G_k / (c_k(j) + 1)$, where $c_k(j)$ is kernel $k$'s resident CTA count on SM $j$ and $G_k$ is its full grid size. This boosts kernels with large remaining grids and decays the score as a kernel accumulates CTAs, preventing any one kernel from monopolizing the SM. (2) \pnamer{} scales $s$ by an inverse register-occupancy factor, $s_{REG} = s / (c_k(j) \cdot R_k + 1)$, where $R_k$ is the per-CTA register count; the denominator is the kernel's current register-file footprint on SM $j$, so register-heavy kernels decay fastest and leave capacity for co-residents.

\noindent\textbf{Comparison Points.}
We compare \pname{} against two baselines: 
(i) \textit{Baseline}, where the VLM of one frame and the action phases of another frame are overlapped by the default GPU scheduler; and (ii) Warped-Slicer~\cite{warpslicer}, the closest prior work, where the VLM and action phases of different frames are overlapped using intra-SM concurrent-kernel scheduling. 
For each kernel, Warped-Slicer~\cite{warpslicer} sweeps the number of resident thread blocks from one to full SM capacity, recording IPC at each CTA count over a 5K-cycle profiling window. The resulting per-kernel IPC-vs-CTA curves are used to assign static per-SM CTA quotas that maximize aggregate IPC.

\section{Evaluation}
\label{sec:eval}

We evaluate \pname{} by measuring the throughput (actions per second) of one VLM phase overlapped with one action phase using four models \cite{cogact, graspvla, pi, octo}. 
We note that the throughput varies across models due to differences in the default action chunk size.

\begin{figure}[h]
\vspace{-0.1cm}
    \centering
    \includegraphics[width=1\linewidth]{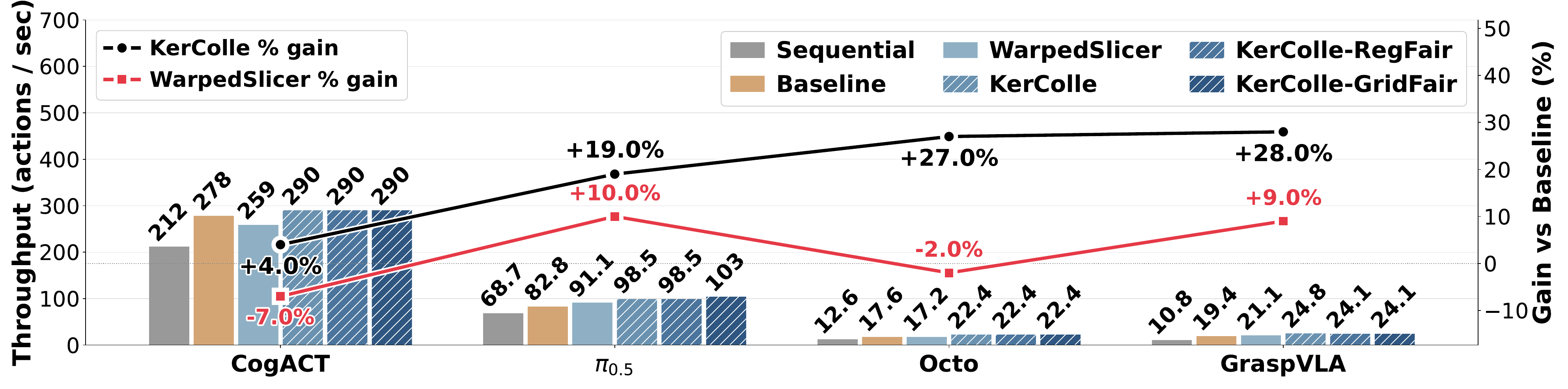}
    \vspace{-0.7cm}
    \caption{Left: VLA inference throughput on RTX 4090. Right: Percentage increase in throughput gain of \pname{} and Warped-Slicer relative to Baseline.}
    \label{fig:speedup4090}
    \vspace{-0.6cm}
\end{figure}
\begin{figure}[h]
    \centering
    \includegraphics[width=1\linewidth]{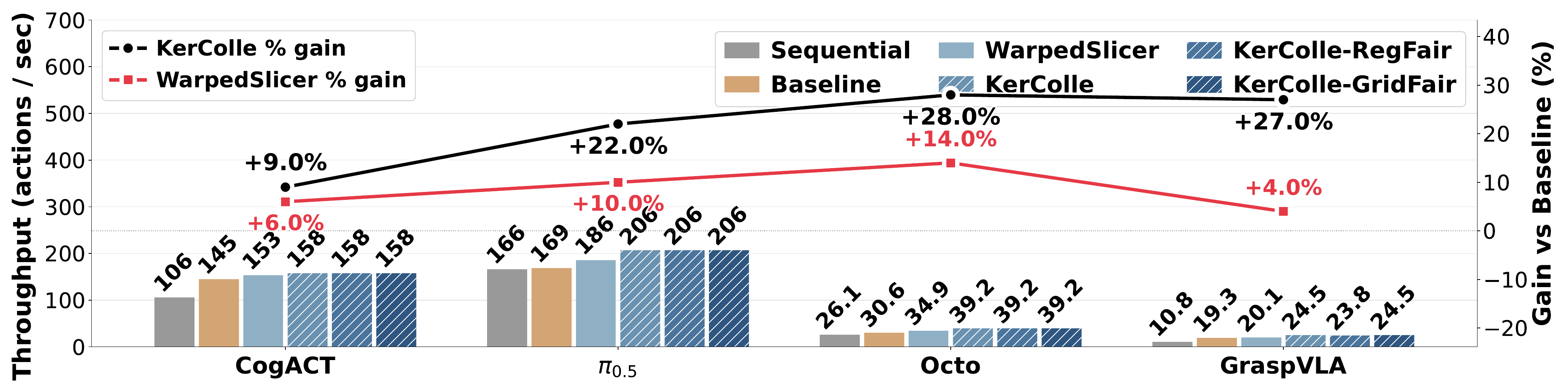}
    \vspace{-0.7cm}
    \caption{Left: VLA inference throughput on RTX 5090. Right: Throughput gain of \pname{} and Warped-Slicer relative to Baseline.}
    \label{fig:speedup5090}
    \vspace{-0.4cm}
\end{figure}

\subsection{Performance Analysis of \pname{}}

Figures~\ref{fig:speedup4090} and~\ref{fig:speedup5090} present the performance benefits of \pname{}, measuring throughput on RTX~4090 and RTX~5090 configurations using Accel-Sim. We make five key observations.

First, we observe that all three variants of \pname{} consistently outperform Warped-Slicer in throughput. The red and black lines indicate the percentage of Warped-Slicer's and \pname{}'s throughput gain relative to the Baseline. The red line is consistently below the black line. This is due to two key reasons: (1) Warped-Slicer's 5K-cycle profiling window is too long for VLA workloads: 18–44\% of kernels, depending on the model, finish in fewer than 5K cycles, forcing Warped-Slicer to fall back to the baseline scheduler for these kernels (Table~\ref{tab:kernel-5k-cycle-cudagraph}). (2) Warped-Slicer's profiling assigns a fixed target CTA count for each SM. This works well for large kernels that can fill the SM, but VLA action phases frequently launch small kernels that cannot reach the profiled target, leaving SMs idle for the entire 5K-cycle window. This problem is worse on the RTX~5090, where the increased SM count results in more idle units, which can cause a slowdown relative to the baseline scheduler.

Second, \pname{}'s throughput gains grow with GPU capability. We observe that \pname{}'s action throughput rises from $19\%$ over baseline on the RTX~4090 to $22\%$ on the RTX~5090 for $\pi_{0.5}$, and from $4\%$ to $9\%$ for CogACT. As the SM count grows from 128 to 170 from the RTX~4090 to the RTX~5090 alongside higher FLOPS and memory bandwidth, individual VLA kernels underutilize the GPU more severely, leaving more SMs idle. This widens the gap that \pname{} can recover through overlap-aware scheduling. The baseline scheduler cannot exploit these idle SMs due to head-of-line blocking, whereas \pname{} fills them with work from the second stream, thus its gains improve further in more powerful GPUs.

Third, CogACT achieves the smallest throughput gain under \pname{} relative to the baseline for two reasons: (1) CogACT achieves the highest concurrent execution under the baseline scheduler, leaving little headroom for \pname{} to improve (Figure~\ref{fig:concurrency}), and (2) the majority of CogACT kernels contain fewer than five thread blocks, limiting the opportunity to co-schedule kernels from both streams onto the same SM (Figure~\ref{fig:cta_asymmetry}). Together, these factors leave \pname{} little room to improve over the baseline.

Fourth, the fairness-augmented variants yield negligible gains over \pname{} in nearly all cases. \pnameg{} matches \pname{} on six of the eight configurations and improves throughput in only one, raising $\pi_{0.5}$ on the RTX~4090 from 98.5 to 103 actions/sec. This is due to $\pi_{0.5}$'s kernel size distribution (Figure~\ref{fig:cta_asymmetry}): $65\%$ of its concurrent kernels contain 11+ CTAs with a median of 30, the highest across all evaluated workloads. These kernels are large enough that the scheduler requires multiple admission decisions to fully schedule them, yet \pname{} may repeatedly prioritize the larger co-running stream, delaying the smaller one. \pnameg{}'s grid-progress factor corrects this by increasing the priority of the delayed kernel as the larger stream accumulates more resident CTAs. This effect does not appear on the RTX~5090, where the additional SM capacity is sufficient to accommodate both streams simultaneously under \pname{}, eliminating any admission contention for \pnameg{} to resolve. 

Finally, \pnamer{} occasionally degrades throughput: GraspVLA drops from $24.5$ to $23.8$ actions/sec on the RTX~5090 and from $24.8$ to $24.1$ on the RTX~4090. This is because \pnamer{} penalizes register-heavy kernels and prioritizes register-light ones instead. However, for GraspVLA, the register-light kernels do not utilize the SM well, so admitting them over the register-heavy ones ultimately wastes SM capacity and reduces throughput.

We conclude that \pname{} consistently improves throughput across diverse VLA models and GPU architectures, with gains growing as GPUs become more powerful. \pname{} alone delivers significant performance improvements, and the fairness-augmented variants achieve comparable gains.

\subsection{Concurrent SM Utilization Analysis}
\label{sec:evalconc}

\begin{figure}[!htb]
    \centering
    \includegraphics[width=1\linewidth]{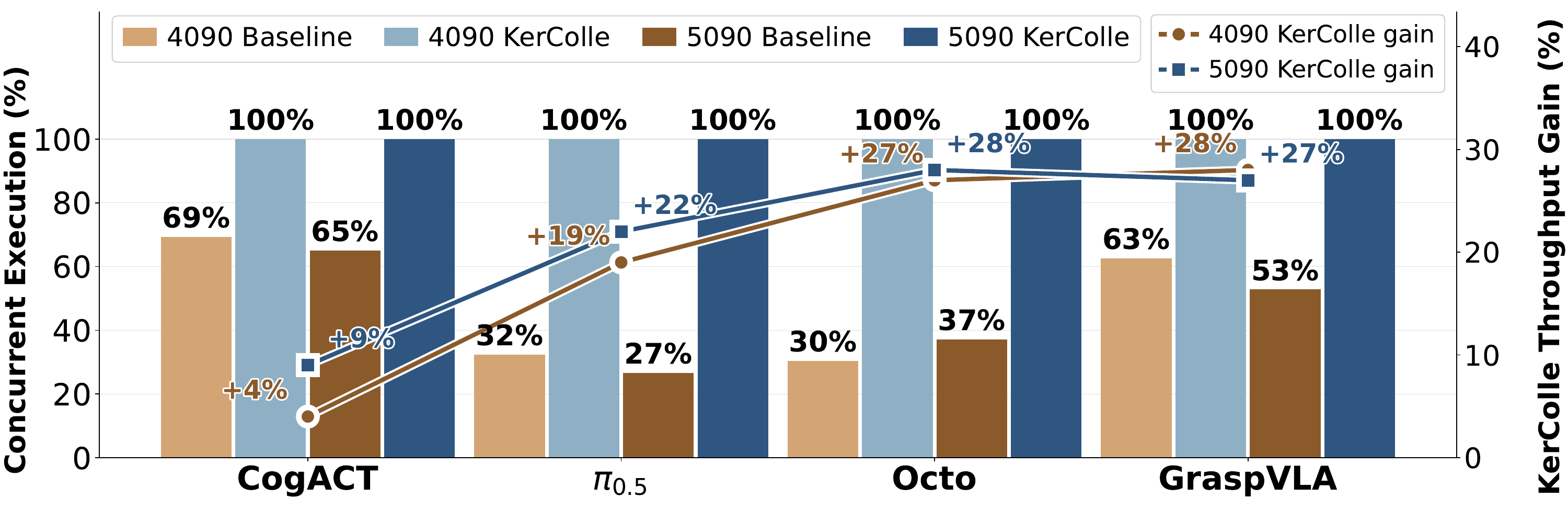}
    \vspace{-0.7cm}
    \caption{Concurrent SM utilization under the baseline scheduler and \pname{} across all models and GPUs (left y-axis), and \pname{} throughput gain relative to baseline (right y-axis).}
    \label{fig:concurrency}
    \vspace{-0.2cm}
\end{figure}

Figure~\ref{fig:concurrency} reports, for each model on both the RTX~4090 and RTX~5090, the fraction of the overlap window during which both streams have at least one thread block resident on every SM, measured under both the baseline scheduler and \pname{} (left y-axis), alongside the corresponding \pname{} throughput gain (right y-axis).

We observe that Octo and GraspVLA achieve the highest throughput gains under \pname{} at 27\%--28\% across both GPUs, while CogACT achieves the lowest at 4\%--9\%. This is because of the differences in concurrency headroom left by the baseline scheduler. Octo achieves only 30--37\% concurrent execution under the baseline scheduler, meaning execution is serial during 63--70\% of the overlap window despite both streams being ready to run. \pname{} converts this serialized execution into true overlap, delivering significant throughput gains. Conversely, CogACT already achieves 65--69\% concurrent execution under the baseline, leaving little room for \pname{} to improve. We conclude that the degree of concurrency left unexploited by the baseline scheduler is the primary determinant of \pname{}'s throughput gains.

\vspace{-0.5em}
\subsection{Effect of Grid-Size Disparity}

\begin{figure}[htbp]
\vspace{-0.1cm}
    \centering
    \includegraphics[width=1\linewidth]{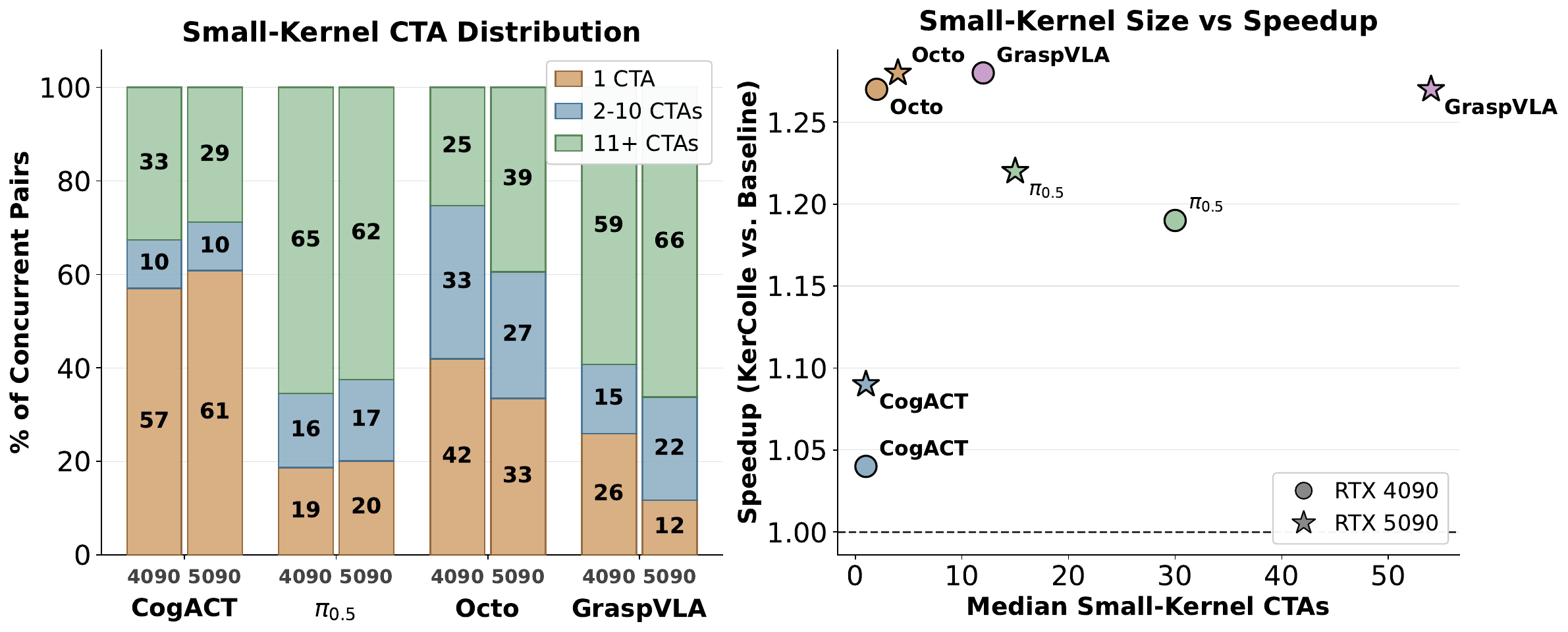}
    \caption{Left:  Distribution of concurrent kernel pairs grouped by the smaller kernel's thread block count. Right:  Throughput gains of \pname{} over baseline for four models correlated to the median thread block count of each model's smaller kernels.}
    \label{fig:cta_asymmetry}
    \vspace{-0.2cm}
\end{figure}

Figure~\ref{fig:cta_asymmetry} analyzes the thread block count of the smaller kernel in each concurrent kernel pair. For each model, the left part of the figure shows the distribution of concurrent kernel pairs grouped by the smaller kernel's thread block count as exactly 1, 2--10, or 11+ thread blocks. The right part of the figure plots each model's throughput gain over baseline, correlated to the median thread block count of its smaller kernels.
Intuitively, the right part of the figure summarizes how small the smaller kernels of concurrent kernel pairs  are for each model, and presents \pname{} throughput gains over baseline scheduler. Grid-size disparity refers to the difference in the number of thread blocks between two kernels that are co-scheduled on the GPU at the same time.

We observe that models whose smaller concurrent kernel contains more thread blocks, such as GraspVLA and $\pi_{0.5}$, experience greater throughput gains under \pname{}. When the smaller kernel in a concurrent pair has very few thread blocks, the grid-size disparity between the two co-scheduled kernels is large, meaning the smaller kernel occupies only a few SMs and overlap is restricted to those SMs alone. In the extreme case of a single-thread-block kernel, the block is pinned to a single SM regardless of the dispatch policy, leaving \pname{} no flexibility to redistribute work across SMs. Conversely, when the smaller kernel has more thread blocks, the grid-size disparity is smaller and both kernels can be interleaved across a larger fraction of SMs, giving the dispatcher more opportunity to overlap them. GraspVLA illustrates the favorable case: only 26\% of its concurrent pairs contain a single-thread-block kernel and its smaller kernels have a median of 11+ thread blocks, allowing the dispatcher to interleave both streams across many SMs and yielding a 28\% throughput gain. 
Although 42\% of Octo's smaller kernels contain only a single thread block, the baseline scheduler left substantial concurrency headroom (Section~\ref{sec:evalconc}), and \pname{} is therefore able to deliver high throughput gains for Octo as well. $\pi_{0.5}$ gains less than GraspVLA despite having ample concurrency headroom and a favorable thread block distribution in its smaller kernels. This is because of the arithmetic intensity of $\pi_{0.5}$'s kernels, as illustrated in Figure~\ref{fig:roofline}: both phases land in the compute-bound region of the roofline model, meaning co-execution finds little complementary headroom  across execution pipelines to exploit.
Overall, we conclude that grid-size disparity between concurrent kernel pairs is a key factor in the overlap capability and throughput gains that \pname{} can achieve.

\subsection{Warp Occupancy Analysis}

\begin{figure}[htb]
    \centering
    \includegraphics[width=1\linewidth]{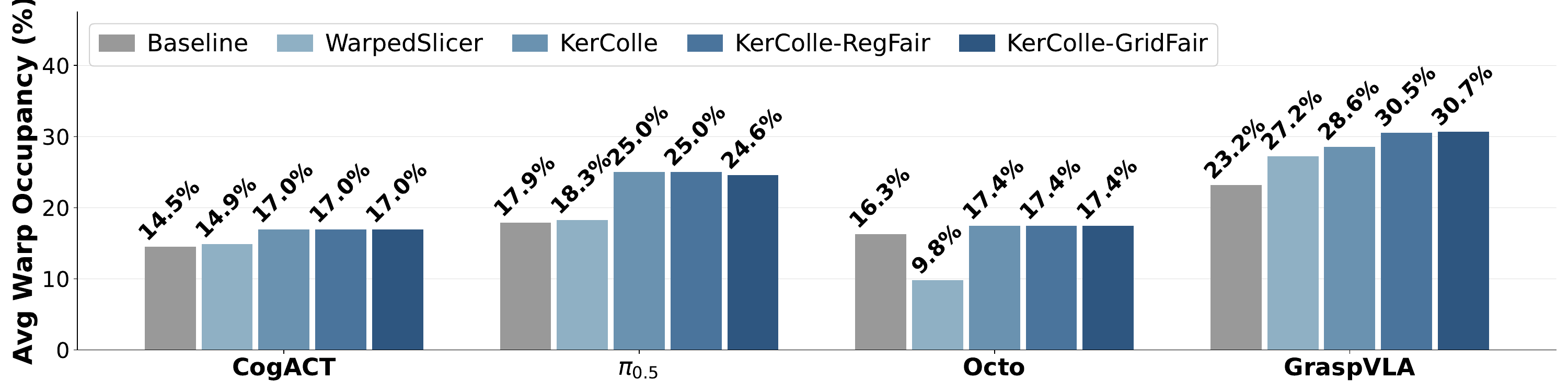}
    \vspace{-0.5cm}
    \caption{Warp occupancy of different \pname{} configurations and Warped-Slicer on RTX~4090.}
    \label{fig:warp_occ_4090}
\vspace{-0.4cm}
\end{figure}

\begin{figure}[htb]
    \centering
    \includegraphics[width=1\linewidth]{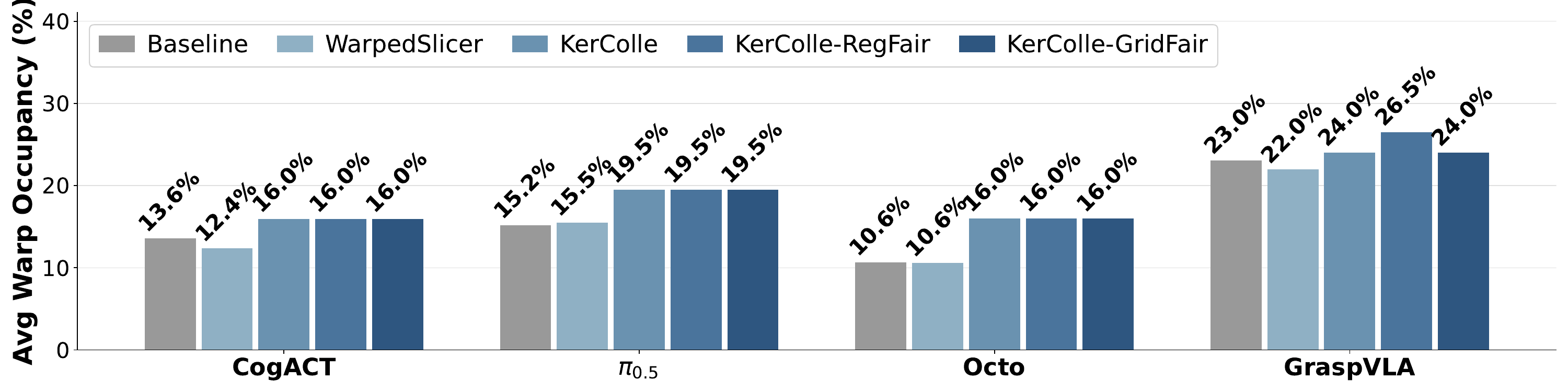}
    \vspace{-0.5cm}
    \caption{Warp occupancy of different \pname{} configurations and Warped-Slicer on RTX~5090.}
    \label{fig:warp_occ_5090}
    \vspace{-0.4cm}

\vspace{-0.1cm}
\end{figure}

Figures~\ref{fig:warp_occ_4090} and~\ref{fig:warp_occ_5090} report average warp occupancy on RTX~4090 and RTX~5090 to show how effectively each scheduler utilizes the per-SM issue capacity.
We observe that \pname{} increases warp occupancy across nearly all models, while Warped-Slicer frequently reduces it below the baseline scheduler.  Specifically, \pname{} improves occupancy over the baseline in seven of eight configurations, with the largest gains occurring where throughput gains are also highest: \pname{} improves warp occupancy in $\pi_{0.5}$  from $17.9\%$ to $25.0\%$ on RTX~4090 and from $15.2\%$ to $19.5\%$ on RTX~5090, and in GraspVLA from $23.2\%$ to $28.6\%$ on RTX~4090 and $23.0\%$ to $24.0\%$ on RTX~5090. These gains confirm that \pname{} improves throughput by filling otherwise-idle SM issue slots with useful work from both streams, rather than simply changing the order in which kernels are dispatched. In contrast, Warped-Slicer frequently \emph{reduces} warp occupancy below the baseline scheduler: Warped-Slicer decreases warp occupancy in Octo from $16.3\%$ to $9.8\%$ on RTX 4090, CogACT from $13.6\%$ to $12.4\%$ on RTX 5090, and GraspVLA from $23.0\%$ to $22.0\%$ on RTX 5090. This is because, under Warped-Slicer, SMs are pinned to a profiling target that they cannot reach, so they hold idle warp slots open for the remaining 5K-cycle window, reducing overall occupancy. 
Overall, we conclude that \pname{} effectively improves warp occupancy across various VLA models and GPUs.


\subsection{Sensitivity to Intra-Kernel CTA Heterogeneity}

When a CTA within the kernel has a resource profile that differs from the majority of CTAs in the same kernel, it will receive the kernel-level intensity tag representing the majority of CTAs and can therefore be misrepresented. Consequently, the scheduler may place the CTA on an SM based on an inaccurate estimate of its resource utilization, resulting in either increased contention or underutilization. To characterize this impact, we inspected 586 unique launch configurations across four VLA models. CTAs differed from their configuration’s typical behavior by at least 10\% in 24.7\% of CogACT configurations, 18.5\% of $\pi0.5$, 11.6\% of Octo, and 22.0\% of GraspVLA. Most cases come from boundary GEMM tiles and copy CTAs performing different roles. Using CogACT on RTX 4090 as a case study, we augmented the simulator to track exact per-CTA profiles during scheduling. Throughput increases only from 290 to 294 actions/s, indicating that CTA-level deviations are too small to change the score $s$ and, consequently, the scheduling decisions. 

\vspace{-0.6em}
\section{Related Work}
\label{sec:related}
To our knowledge, \pname{} is the first work to leave closed-source VLM kernels unmodified and augment the thread block dispatcher to co-locate thread blocks from kernels exhibiting different functional-unit demands while mitigating head-of-line blocking.
Prior approaches to GPU kernel co-location each have limitations for VLA inference. 

\textbf{Preemption and Runtime Interposition.}
Hardware preemption~\cite{park2015chimera, tanasic2014srtf, wu2017flep, Maestro, smk} is too coarse for sub-microsecond kernels. Runtime interposition~\cite{tally, orion, reef, paella, chen2016baymax, chen2017prophet} adds per-launch overhead that short action-head kernels cannot amortize. Persistent-block transformations~\cite{pai2013elastic, allen2019slate, liang2020modelbasedsmk, zhao2021plasticine, ispa} and kernel fusion~\cite{li2022hfuse, tacker, aker} require source code access. Thread block scheduler modifications~\cite{kim2019dynsharing, awatramani2013kits, lin2019coordinated, shekofteh2020ccuda, kim2021kscheduler, atb, wu2015smcentric} and intra-kernel auxiliary work~\cite{vijaykumar2015caba, kayiran2016ucstates, latoa, ho} either fail to eliminate head-of-line blocking or operate within a single kernel.

\textbf{Spatial Partitioning.}
Prior approaches in spatial partitioning~\cite{mps, mig, adriaens2012spatial, aguilera2014fairshare, aguilera2014qos, liang2016kernelmgmt, zhao2018classification} allocate disjoint sets of SMs to concurrently execute kernels. While this approach mitigates head-of-line blocking, it is ineffective for VLAs since a VLA's two phases both underutilize their assigned SMs, so partitioning the SMs into kernel-exclusive sets does not mitigate the underutilization. For VLAs, the goal is to co-locate the kernels from distinct phases onto the same SM to exploit the different resource demands.

\textbf{Data- and Locality-Aware Scheduling.}
Prior approaches explore data- and cache-aware intra-SM warp schedulers~\cite{rogers2012ccws, rogers2013daws, narasiman2011lwm, lee2014caws, liu2015saws, gpucke}, locality-aware CTA scheduling that captures inter-kernel and inter-CTA reuse~\cite{flipsteal, paver, cluster, wang2016laperm, chen2017tbs, belviranli2016cumas}, and CTA dispatchers that honor inter-block DAG dependencies~\cite{blockmaestro, wireframe}. However, these lines of work target cache contention within a single application or thread-block data dependencies—neither of which is the primary bottleneck in VLA workloads—and do not address head-of-line blocking or co-execute thread blocks from kernels with complementary execution-unit demands.



\vspace{-0.15cm}



\section{Conclusion}
We characterized an important emerging workload for GPUs in Physical AI and we identified a gap in existing GPU scheduling frameworks: applications whose set of concurrently running kernels shifts continuously render static or ahead-of-time scheduling policies largely ineffective; applications dominated by short-running, numerous kernels leave no headroom for online profiling or preemption-based mechanisms. To address these challenges, we propose a lightweight scheduling framework that leverages online SM utilization and individual kernel resource requirements to intelligently and dynamically co-schedule kernels to efficiently overlap execution. We demonstrate that \pname{} improves four representative VLA models' throughputs by enabling more efficient concurrent execution of their independent heterogeneous phases. 

\bibliographystyle{ACM-Reference-Format}
\bibliography{sample-base}

@misc{gilman_sigmetrics,
      title={Characterizing Concurrency Mechanisms for NVIDIA GPUs under Deep Learning Workloads}, 
      author={Guin Gilman and Robert J. Walls},
      year={2021},
      eprint={2110.00459},
      archivePrefix={arXiv},
      primaryClass={cs.DC},
      url={https://arxiv.org/abs/2110.00459}, 
}

@INPROCEEDINGS{hardwarep,
  author={Bakita, Joshua and Anderson, James H.},
  booktitle={2023 IEEE 29th Real-Time and Embedded Technology and Applications Symposium (RTAS)}, 
  title={Hardware Compute Partitioning on NVIDIA GPUs}, 
  year={2023},
  volume={},
  number={},
  pages={54-66},
  doi={10.1109/RTAS58335.2023.00012}}

@inproceedings{tally,
    author      = {Zhao, Wei and Jayarajan, Anand and Pekhimenko, Gennady},
    title       = {Tally: Non-Intrusive Performance Isolation for Concurrent Deep Learning Workloads},
    year        = {2025},
    url         = {https://doi.org/10.1145/3669940.3707282},
    doi         = {10.1145/3669940.3707282},
    booktitle   = {Proceedings of the 30th ACM International Conference on Architectural Support for Programming Languages and Operating Systems, Volume 1},
    pages       = {1052–1068},
    series      = {ASPLOS '25}
}

@misc{kitsune,
      title={Kitsune: Enabling Dataflow Execution on GPUs}, 
      author={Michael Davies and Neal Crago and Karthikeyan Sankaralingam and Stephen W. Keckler},
      year={2025},
      eprint={2502.18403},
      archivePrefix={arXiv},
      primaryClass={cs.AR},
      url={https://arxiv.org/abs/2502.18403}, 
}

@inproceedings{gpupool,
    author = {Tan, Xiaodan Serina and Golikov, Pavel and Vijaykumar, Nandita and Pekhimenko, Gennady},
    title = {GPUPool: A Holistic Approach to Fine-Grained GPU Sharing in the Cloud},
    year = {2023},
    isbn = {9781450398688},
    publisher = {Association for Computing Machinery},
    address = {New York, NY, USA},
    url = {https://doi.org/10.1145/3559009.3569650},
    doi = {10.1145/3559009.3569650},
    booktitle = {Proceedings of the International Conference on Parallel Architectures and Compilation Techniques},
    pages = {317–332},
    numpages = {16},
    location = {Chicago, Illinois},
    series = {PACT '22}
}

@inproceedings{warpslicer,
author = {Xu, Qiumin and Jeon, Hyeran and Kim, Keunsoo and Ro, Won Woo and Annavaram, Murali},
title = {Warped-slicer: efficient intra-SM slicing through dynamic resource partitioning for GPU multiprogramming},
year = {2016},
isbn = {9781467389471},
publisher = {IEEE Press},
url = {https://doi.org/10.1109/ISCA.2016.29},
doi = {10.1109/ISCA.2016.29},
booktitle = {Proceedings of the 43rd International Symposium on Computer Architecture},
pages = {230–242},
numpages = {13},
location = {Seoul, Republic of Korea},
series = {ISCA '16}
}

@inproceedings{smk,
  author={Wang, Zhenning and Yang, Jun and Melhem, Rami and Childers, Bruce and Zhang, Youtao and Guo, Minyi},
  booktitle={2016 IEEE International Symposium on High Performance Computer Architecture (HPCA)}, 
  title={Simultaneous Multikernel GPU: Multi-tasking throughput processors via fine-grained sharing}, 
  year={2016},
  volume={},
  number={},
  pages={358-369},
  doi={10.1109/HPCA.2016.7446078}}

@ARTICLE{ispa,
  author={Zhao, Han and Cui, Weihao and Chen, Quan and Guo, Minyi},
  journal={IEEE Transactions on Computers}, 
  title={ISPA: Exploiting Intra-SM Parallelism in GPUs via Fine-Grained Resource Management}, 
  year={2023},
  volume={72},
  number={5},
  pages={1473-1487},
  doi={10.1109/TC.2022.3214088}}

@INPROCEEDINGS{ho,
  author={Ho, Khoa and Zhao, Hui and Jog, Adwait and Mohanty, Saraju},
  booktitle={2022 IEEE Computer Society Annual Symposium on VLSI (ISVLSI)}, 
  title={Improving GPU Throughput through Parallel Execution Using Tensor Cores and CUDA Cores}, 
  year={2022},
  volume={},
  number={},
  pages={223-228},
  doi={10.1109/ISVLSI54635.2022.00051}}

@INPROCEEDINGS{tacker,
  author={Zhao, Han and Cui, Weihao and Chen, Quan and Zhang, Youtao and Lu, Yanchao and Li, Chao and Leng, Jingwen and Guo, Minyi},
  booktitle={2022 IEEE International Symposium on High-Performance Computer Architecture (HPCA)}, 
  title={Tacker: Tensor-CUDA Core Kernel Fusion for Improving the GPU Utilization while Ensuring QoS}, 
  year={2022},
  volume={},
  number={},
  pages={800-813},
  doi={10.1109/HPCA53966.2022.00064}}

@inproceedings{Maestro,
author = {Park, Jason Jong Kyu and Park, Yongjun and Mahlke, Scott},
title = {Dynamic Resource Management for Efficient Utilization of Multitasking GPUs},
year = {2017},
isbn = {9781450344654},
publisher = {Association for Computing Machinery},
address = {New York, NY, USA},
url = {https://doi.org/10.1145/3037697.3037707},
doi = {10.1145/3037697.3037707},
booktitle = {Proceedings of the Twenty-Second International Conference on Architectural Support for Programming Languages and Operating Systems},
pages = {527–540},
numpages = {14},
location = {Xi'an, China},
series = {ASPLOS '17}
}

@INPROCEEDINGS{atb,
  author={Lee, Minseok and Song, Seokwoo and Moon, Joosik and Kim, John and Seo, Woong and Cho, Yeongon and Ryu, Soojung},
  booktitle={2014 IEEE 20th International Symposium on High Performance Computer Architecture (HPCA)}, 
  title={Improving GPGPU resource utilization through alternative thread block scheduling}, 
  year={2014},
  volume={},
  number={},
  pages={260-271},
  doi={10.1109/HPCA.2014.6835937}}

@article{aker,
author = {Zhao, Han and Deng, Junxiao and Cui, Weihao and Chen, Quan and Zhang, Youtao and Zeng, Deze and Guo, Minyi},
title = {Adaptive Kernel Fusion for Improving the GPU Utilization While Ensuring QoS},
year = {2025},
issue_date = {Feb. 2025},
publisher = {IEEE Computer Society},
address = {USA},
volume = {74},
number = {2},
issn = {0018-9340},
url = {https://doi.org/10.1109/TC.2024.3477995},
doi = {10.1109/TC.2024.3477995},
journal = {IEEE Trans. Comput.},
month = feb,
pages = {386–400},
numpages = {15}
}

@inproceedings{paella,
author = {Ng, Kelvin K. W. and Demoulin, Henri Maxime and Liu, Vincent},
title = {Paella: Low-latency Model Serving with Software-defined GPU Scheduling},
year = {2023},
isbn = {9798400702297},
publisher = {Association for Computing Machinery},
address = {New York, NY, USA},
url = {https://doi.org/10.1145/3600006.3613163},
doi = {10.1145/3600006.3613163},
booktitle = {Proceedings of the 29th Symposium on Operating Systems Principles},
pages = {595–610},
numpages = {16},
location = {Koblenz, Germany},
series = {SOSP '23}
}

@inproceedings {reef,
author = {Mingcong Han and Hanze Zhang and Rong Chen and Haibo Chen},
title = {Microsecond-scale Preemption for Concurrent {GPU-accelerated} {DNN} Inferences},
booktitle = {16th USENIX Symposium on Operating Systems Design and Implementation (OSDI 22)},
year = {2022},
isbn = {978-1-939133-28-1},
address = {Carlsbad, CA},
pages = {539--558},
url = {https://www.usenix.org/conference/osdi22/presentation/han},
publisher = {USENIX Association},
month = jul
}

@inproceedings{orion,
author = {Strati, Foteini and Ma, Xianzhe and Klimovic, Ana},
title = {Orion: Interference-aware, Fine-grained GPU Sharing for ML Applications},
year = {2024},
isbn = {9798400704376},
publisher = {Association for Computing Machinery},
address = {New York, NY, USA},
url = {https://doi.org/10.1145/3627703.3629578},
doi = {10.1145/3627703.3629578},
booktitle = {Proceedings of the Nineteenth European Conference on Computer Systems},
pages = {1075–1092},
numpages = {18},
location = {Athens, Greece},
series = {EuroSys '24}
}

@misc{mps,
  author = {{NVIDIA Corporation}},
  title  = {{Multi-Process Service (MPS)}},
  year   = {2024},
  url    = {https://docs.nvidia.com/deploy/mps/},
  note   = {Accessed: 2025}
}

@misc{mig,
  author = {{NVIDIA Corporation}},
  title  = {{NVIDIA Multi-Instance GPU User Guide}},
  year   = {2024},
  url    = {https://docs.nvidia.com/datacenter/tesla/mig-user-guide/},
  note   = {Accessed: 2025}
}

@manual{ncu,
  title  = {{NVIDIA Nsight Compute} -- Kernel Profiling Guide},
  author = {{NVIDIA Corporation}},
  year   = {2024},
  url    = {https://docs.nvidia.com/nsight-compute/ProfilingGuide/}
}

@misc{nvidia_ada_whitepaper,
  title   = {{NVIDIA Ada GPU Architecture}},
  author  = {NVIDIA},
  year    = {2022},
  howpublished = {\url{https://images.nvidia.com/aem-dam/Solutions/geforce/ada/nvidia-ada-gpu-architecture.pdf}},
  note    = {Whitepaper v2.02}
}

@techreport{nvblackwell,
  author       = {{NVIDIA Corporation}},
  title        = {{NVIDIA RTX Blackwell GPU Architecture: Built for Neural Rendering}},
  institution  = {NVIDIA Corporation},
  year         = {2025},
  type         = {Whitepaper},
  number       = {V1.1},
  url          = {https://images.nvidia.com/aem-dam/Solutions/geforce/blackwell/nvidia-rtx-blackwell-gpu-architecture.pdf},
  note         = {Accessed: 2026-04-16}
}

@article{cudnn,
  title={cuDNN: Efficient Primitives for Deep Learning},
  author={Chetlur, Sharan and Woolley, Cliff and Vandermersch, Philippe and Cohen, Jonathan and Tran, John and Catanzaro, Bryan and Shelhamer, Evan},
  journal={arXiv preprint arXiv:1410.0759},
  year={2014}
}

@misc{cublas,
  author = {{NVIDIA Corporation}},
  title = {{cuBLAS Library}},
  howpublished = {\url{https://docs.nvidia.com/cuda/cublas/}},
  note = {Accessed: 2026-04-22}
}

@inproceedings{nvbit,
author = {Villa, Oreste and Stephenson, Mark and Nellans, David and Keckler, Stephen W.},
title = {NVBit: A Dynamic Binary Instrumentation Framework for NVIDIA GPUs},
year = {2019},
isbn = {9781450369381},
publisher = {Association for Computing Machinery},
address = {New York, NY, USA},
url = {https://doi.org/10.1145/3352460.3358307},
doi = {10.1145/3352460.3358307},
booktitle = {Proceedings of the 52nd Annual IEEE/ACM International Symposium on Microarchitecture},
pages = {372–383},
numpages = {12},
location = {Columbus, OH, USA},
series = {MICRO-52}
}

@manual{cudaapi,
  title        = {{CUDA C++ Programming Guide}},
  author       = {{NVIDIA Corporation}},
  year         = {2025},
  note         = {Version 12.6},
  url          = {https://docs.nvidia.com/cuda/cuda-c-programming-guide/},
}

@INPROCEEDINGS{accelsim,
  author={Khairy, Mahmoud and Shen, Zhesheng and Aamodt, Tor M. and Rogers, Timothy G.},
  booktitle={2020 ACM/IEEE 47th Annual International Symposium on Computer Architecture (ISCA)}, 
  title={Accel-Sim: An Extensible Simulation Framework for Validated GPU Modeling}, 
  year={2020},
  volume={},
  number={},
  pages={473-486},
  doi={10.1109/ISCA45697.2020.00047}}

@misc{dynamicvla,
      title={DynamicVLA: A Vision-Language-Action Model for Dynamic Object Manipulation}, 
      author={Haozhe Xie and Beichen Wen and Jiarui Zheng and Zhaoxi Chen and Fangzhou Hong and Haiwen Diao and Ziwei Liu},
      year={2026},
      eprint={2601.22153},
      archivePrefix={arXiv},
      primaryClass={cs.RO},
      url={https://arxiv.org/abs/2601.22153}, 
}

@misc{adaptive_chunking,
      title={Adaptive Action Chunking at Inference-time for Vision-Language-Action Models}, 
      author={Yuanchang Liang and Xiaobo Wang and Kai Wang and Shuo Wang and Xiaojiang Peng and Haoyu Chen and David Kim Huat Chua and Prahlad Vadakkepat},
      year={2026},
      eprint={2604.04161},
      archivePrefix={arXiv},
      primaryClass={cs.RO},
      url={https://arxiv.org/abs/2604.04161}, 
}

@article{openvla,
    title={OpenVLA: An Open-Source Vision-Language-Action Model},
    author={{Moo Jin} Kim and Karl Pertsch and Siddharth Karamcheti and Ted Xiao and Ashwin Balakrishna and Suraj Nair and Rafael Rafailov and Ethan Foster and Grace Lam and Pannag Sanketi and Quan Vuong and Thomas Kollar and Benjamin Burchfiel and Russ Tedrake and Dorsa Sadigh and Sergey Levine and Percy Liang and Chelsea Finn},
    journal = {arXiv preprint arXiv:2406.09246},
    year={2024},
}

@misc{efficient_vla_survey,
      title={A Survey on Efficient Vision-Language-Action Models}, 
      author={Zhaoshu Yu and Bo Wang and Pengpeng Zeng and Haonan Zhang and Ji Zhang and Zheng Wang and Lianli Gao and Jingkuan Song and Nicu Sebe and Heng Tao Shen},
      year={2026},
      eprint={2510.24795},
      archivePrefix={arXiv},
      primaryClass={cs.CV},
      url={https://arxiv.org/abs/2510.24795}, 
}

@misc{vlaperf,
      title={How Fast Can I Run My VLA? Demystifying VLA Inference Performance with VLA-Perf}, 
      author={Wenqi Jiang and Jason Clemons and Karu Sankaralingam and Christos Kozyrakis},
      year={2026},
      eprint={2602.18397},
      archivePrefix={arXiv},
      primaryClass={cs.RO},
      url={https://arxiv.org/abs/2602.18397}, 
}

@misc{tinyvla,
      title={TinyVLA: Towards Fast, Data-Efficient Vision-Language-Action Models for Robotic Manipulation}, 
      author={Junjie Wen and Yichen Zhu and Jinming Li and Minjie Zhu and Kun Wu and Zhiyuan Xu and Ning Liu and Ran Cheng and Chaomin Shen and Yaxin Peng and Feifei Feng and Jian Tang},
      year={2025},
      eprint={2409.12514},
      archivePrefix={arXiv},
      primaryClass={cs.RO},
      url={https://arxiv.org/abs/2409.12514}, 
}

@misc{openvla_oft,
      title={Fine-Tuning Vision-Language-Action Models: Optimizing Speed and Success}, 
      author={Moo Jin Kim and Chelsea Finn and Percy Liang},
      year={2025},
      eprint={2502.19645},
      archivePrefix={arXiv},
      primaryClass={cs.RO},
      url={https://arxiv.org/abs/2502.19645}, 
}

@misc{duocore,
      title={Asynchronous Fast-Slow Vision-Language-Action Policies for Whole-Body Robotic Manipulation}, 
      author={Teqiang Zou and Hongliang Zeng and Yuxuan Nong and Yifan Li and Kehui Liu and Haotian Yang and Xinyang Ling and Xin Li and Lianyang Ma},
      year={2025},
      eprint={2512.20188},
      archivePrefix={arXiv},
      primaryClass={cs.RO},
      url={https://arxiv.org/abs/2512.20188}, 
}

@misc{graspvla,
      title={GraspVLA: a Grasping Foundation Model Pre-trained on Billion-scale Synthetic Action Data}, 
      author={Shengliang Deng and Mi Yan and Songlin Wei and Haixin Ma and Yuxin Yang and Jiayi Chen and Zhiqi Zhang and Taoyu Yang and Xuheng Zhang and Wenhao Zhang and Heming Cui and Zhizheng Zhang and He Wang},
      year={2025},
      eprint={2505.03233},
      archivePrefix={arXiv},
      primaryClass={cs.RO},
      url={https://arxiv.org/abs/2505.03233}, 
}

@misc{octo,
      title={Octo: An Open-Source Generalist Robot Policy}, 
      author={Octo Model Team and Dibya Ghosh and Homer Walke and Karl Pertsch and Kevin Black and Oier Mees and Sudeep Dasari and Joey Hejna and Tobias Kreiman and Charles Xu and Jianlan Luo and You Liang Tan and Lawrence Yunliang Chen and Pannag Sanketi and Quan Vuong and Ted Xiao and Dorsa Sadigh and Chelsea Finn and Sergey Levine},
      year={2024},
      eprint={2405.12213},
      archivePrefix={arXiv},
      primaryClass={cs.RO},
      url={https://arxiv.org/abs/2405.12213}, 
}

@misc{vlacache,
      title={VLA-Cache: Efficient Vision-Language-Action Manipulation via Adaptive Token Caching}, 
      author={Siyu Xu and Yunke Wang and Chenghao Xia and Dihao Zhu and Tao Huang and Chang Xu},
      year={2025},
      eprint={2502.02175},
      archivePrefix={arXiv},
      primaryClass={cs.RO},
      url={https://arxiv.org/abs/2502.02175}, 
}

@misc{realtime_speed,
      title={Running VLAs at Real-time Speed}, 
      author={Yunchao Ma and Yizhuang Zhou and Yunhuan Yang and Tiancai Wang and Haoqiang Fan},
      year={2025},
      eprint={2510.26742},
      archivePrefix={arXiv},
      primaryClass={cs.RO},
      url={https://arxiv.org/abs/2510.26742}, 
}

@misc{vlash,
      title={VLASH: Real-Time VLAs via Future-State-Aware Asynchronous Inference}, 
      author={Jiaming Tang and Yufei Sun and Yilong Zhao and Shang Yang and Yujun Lin and Zhuoyang Zhang and James Hou and Yao Lu and Zhijian Liu and Song Han},
      year={2025},
      eprint={2512.01031},
      archivePrefix={arXiv},
      primaryClass={cs.RO},
      url={https://arxiv.org/abs/2512.01031}, 
}

@misc{faster,
      title={FASTER: Rethinking Real-Time Flow VLAs}, 
      author={Yuxiang Lu and Zhe Liu and Xianzhe Fan and Zhenya Yang and Jinghua Hou and Junyi Li and Kaixin Ding and Hengshuang Zhao},
      year={2026},
      eprint={2603.19199},
      archivePrefix={arXiv},
      primaryClass={cs.RO},
      url={https://arxiv.org/abs/2603.19199}, 
}

@misc{univla,
      title={Unified Vision-Language-Action Model}, 
      author={Yuqi Wang and Xinghang Li and Wenxuan Wang and Junbo Zhang and Yingyan Li and Yuntao Chen and Xinlong Wang and Zhaoxiang Zhang},
      year={2025},
      eprint={2506.19850},
      archivePrefix={arXiv},
      primaryClass={cs.CV},
      url={https://arxiv.org/abs/2506.19850}, 
}

@misc{rtc,
      title={Real-Time Execution of Action Chunking Flow Policies}, 
      author={Kevin Black and Manuel Y. Galliker and Sergey Levine},
      year={2025},
      eprint={2506.07339},
      archivePrefix={arXiv},
      primaryClass={cs.RO},
      url={https://arxiv.org/abs/2506.07339}, 
}

@misc{nvgr001,
      title={GR00T N1: An Open Foundation Model for Generalist Humanoid Robots}, 
      author={NVIDIA and : and Johan Bjorck and Fernando Castañeda and Nikita Cherniadev and Xingye Da and Runyu Ding and Linxi "Jim" Fan and Yu Fang and Dieter Fox and Fengyuan Hu and Spencer Huang and Joel Jang and Zhenyu Jiang and Jan Kautz and Kaushil Kundalia and Lawrence Lao and Zhiqi Li and Zongyu Lin and Kevin Lin and Guilin Liu and Edith Llontop and Loic Magne and Ajay Mandlekar and Avnish Narayan and Soroush Nasiriany and Scott Reed and You Liang Tan and Guanzhi Wang and Zu Wang and Jing Wang and Qi Wang and Jiannan Xiang and Yuqi Xie and Yinzhen Xu and Zhenjia Xu and Seonghyeon Ye and Zhiding Yu and Ao Zhang and Hao Zhang and Yizhou Zhao and Ruijie Zheng and Yuke Zhu},
      year={2025},
      eprint={2503.14734},
      archivePrefix={arXiv},
      primaryClass={cs.RO},
      url={https://arxiv.org/abs/2503.14734}, 
}

@misc{cogact,
      title={CogACT: A Foundational Vision-Language-Action Model for Synergizing Cognition and Action in Robotic Manipulation}, 
      author={Qixiu Li and Yaobo Liang and Zeyu Wang and Lin Luo and Xi Chen and Mozheng Liao and Fangyun Wei and Yu Deng and Sicheng Xu and Yizhong Zhang and Xiaofan Wang and Bei Liu and Jianlong Fu and Jianmin Bao and Dong Chen and Yuanchun Shi and Jiaolong Yang and Baining Guo},
      year={2024},
      eprint={2411.19650},
      archivePrefix={arXiv},
      primaryClass={cs.RO},
      url={https://arxiv.org/abs/2411.19650}, 
}

@misc{fis,
      title={Fast-in-Slow: A Dual-System Foundation Model Unifying Fast Manipulation within Slow Reasoning}, 
      author={Hao Chen and Jiaming Liu and Chenyang Gu and Zhuoyang Liu and Renrui Zhang and Xiaoqi Li and Xiao He and Yandong Guo and Chi-Wing Fu and Shanghang Zhang and Pheng-Ann Heng},
      year={2025},
      eprint={2506.01953},
      archivePrefix={arXiv},
      primaryClass={cs.RO},
      url={https://arxiv.org/abs/2506.01953}, 
}

@misc{smolvla,
      title={SmolVLA: A Vision-Language-Action Model for Affordable and Efficient Robotics}, 
      author={Mustafa Shukor and Dana Aubakirova and Francesco Capuano and Pepijn Kooijmans and Steven Palma and Adil Zouitine and Michel Aractingi and Caroline Pascal and Martino Russi and Andres Marafioti and Simon Alibert and Matthieu Cord and Thomas Wolf and Remi Cadene},
      year={2025},
      eprint={2506.01844},
      archivePrefix={arXiv},
      primaryClass={cs.LG},
      url={https://arxiv.org/abs/2506.01844}, 
}

@misc{pi,
      title={$\pi_{0.5}$: a Vision-Language-Action Model with Open-World Generalization}, 
      author={Physical Intelligence and Kevin Black and Noah Brown and James Darpinian and Karan Dhabalia and Danny Driess and Adnan Esmail and Michael Equi and Chelsea Finn and Niccolo Fusai and Manuel Y. Galliker and Dibya Ghosh and Lachy Groom and Karol Hausman and Brian Ichter and Szymon Jakubczak and Tim Jones and Liyiming Ke and Devin LeBlanc and Sergey Levine and Adrian Li-Bell and Mohith Mothukuri and Suraj Nair and Karl Pertsch and Allen Z. Ren and Lucy Xiaoyang Shi and Laura Smith and Jost Tobias Springenberg and Kyle Stachowicz and James Tanner and Quan Vuong and Homer Walke and Anna Walling and Haohuan Wang and Lili Yu and Ury Zhilinsky},
      year={2025},
      eprint={2504.16054},
      archivePrefix={arXiv},
      primaryClass={cs.LG},
      url={https://arxiv.org/abs/2504.16054}, 
}

@misc{paligemma,
      title={PaliGemma: A versatile 3B VLM for transfer}, 
      author={Lucas Beyer and Andreas Steiner and André Susano Pinto and Alexander Kolesnikov and Xiao Wang and Daniel Salz and Maxim Neumann and Ibrahim Alabdulmohsin and Michael Tschannen and Emanuele Bugliarello and Thomas Unterthiner and Daniel Keysers and Skanda Koppula and Fangyu Liu and Adam Grycner and Alexey Gritsenko and Neil Houlsby and Manoj Kumar and Keran Rong and Julian Eisenschlos and Rishabh Kabra and Matthias Bauer and Matko Bošnjak and Xi Chen and Matthias Minderer and Paul Voigtlaender and Ioana Bica and Ivana Balazevic and Joan Puigcerver and Pinelopi Papalampidi and Olivier Henaff and Xi Xiong and Radu Soricut and Jeremiah Harmsen and Xiaohua Zhai},
      year={2024},
      eprint={2407.07726},
      archivePrefix={arXiv},
      primaryClass={cs.CV},
      url={https://arxiv.org/abs/2407.07726}, 
}

@misc{prismatic,
      title={Prismatic VLMs: Investigating the Design Space of Visually-Conditioned Language Models}, 
      author={Siddharth Karamcheti and Suraj Nair and Ashwin Balakrishna and Percy Liang and Thomas Kollar and Dorsa Sadigh},
      year={2024},
      eprint={2402.07865},
      archivePrefix={arXiv},
      primaryClass={cs.CV},
      url={https://arxiv.org/abs/2402.07865}, 
}

@misc{droid,
      title={DROID: A Large-Scale In-The-Wild Robot Manipulation Dataset}, 
      author={Alexander Khazatsky and Karl Pertsch and Suraj Nair and Ashwin Balakrishna and Sudeep Dasari and Siddharth Karamcheti and Soroush Nasiriany and Mohan Kumar Srirama and Lawrence Yunliang Chen and Kirsty Ellis and Peter David Fagan and Joey Hejna and Masha Itkina and Marion Lepert and Yecheng Jason Ma and Patrick Tree Miller and Jimmy Wu and Suneel Belkhale and Shivin Dass and Huy Ha and Arhan Jain and Abraham Lee and Youngwoon Lee and Marius Memmel and Sungjae Park and Ilija Radosavovic and Kaiyuan Wang and Albert Zhan and Kevin Black and Cheng Chi and Kyle Beltran Hatch and Shan Lin and Jingpei Lu and Jean Mercat and Abdul Rehman and Pannag R Sanketi and Archit Sharma and Cody Simpson and Quan Vuong and Homer Rich Walke and Blake Wulfe and Ted Xiao and Jonathan Heewon Yang and Arefeh Yavary and Tony Z. Zhao and Christopher Agia and Rohan Baijal and Mateo Guaman Castro and Daphne Chen and Qiuyu Chen and Trinity Chung and Jaimyn Drake and Ethan Paul Foster and Jensen Gao and Vitor Guizilini and David Antonio Herrera and Minho Heo and Kyle Hsu and Jiaheng Hu and Muhammad Zubair Irshad and Donovon Jackson and Charlotte Le and Yunshuang Li and Kevin Lin and Roy Lin and Zehan Ma and Abhiram Maddukuri and Suvir Mirchandani and Daniel Morton and Tony Nguyen and Abigail O'Neill and Rosario Scalise and Derick Seale and Victor Son and Stephen Tian and Emi Tran and Andrew E. Wang and Yilin Wu and Annie Xie and Jingyun Yang and Patrick Yin and Yunchu Zhang and Osbert Bastani and Glen Berseth and Jeannette Bohg and Ken Goldberg and Abhinav Gupta and Abhishek Gupta and Dinesh Jayaraman and Joseph J Lim and Jitendra Malik and Roberto Martín-Martín and Subramanian Ramamoorthy and Dorsa Sadigh and Shuran Song and Jiajun Wu and Michael C. Yip and Yuke Zhu and Thomas Kollar and Sergey Levine and Chelsea Finn},
      year={2025},
      eprint={2403.12945},
      archivePrefix={arXiv},
      primaryClass={cs.RO},
      url={https://arxiv.org/abs/2403.12945}, 
}

@article{kawaharazuka_survey,
   title={Vision-Language-Action Models for Robotics: A Review Towards Real-World Applications},
   volume={13},
   ISSN={2169-3536},
   url={http://dx.doi.org/10.1109/ACCESS.2025.3609980},
   DOI={10.1109/access.2025.3609980},
   journal={IEEE Access},
   publisher={Institute of Electrical and Electronics Engineers (IEEE)},
   author={Kawaharazuka, Kento and Oh, Jihoon and Yamada, Jun and Posner, Ingmar and Zhu, Yuke},
   year={2025},
   pages={162467–162504} }

@misc{pure_vla_survey,
      title={Pure Vision Language Action (VLA) Models: A Comprehensive Survey}, 
      author={Dapeng Zhang and Jing Sun and Chenghui Hu and Xiaoyan Wu and Zhenlong Yuan and Rui Zhou and Fei Shen and Qingguo Zhou},
      year={2025},
      eprint={2509.19012},
      archivePrefix={arXiv},
      primaryClass={cs.RO},
      url={https://arxiv.org/abs/2509.19012}, 
}

@misc{spatialvla,
      title={SpatialVLA: Exploring Spatial Representations for Visual-Language-Action Model}, 
      author={Delin Qu and Haoming Song and Qizhi Chen and Yuanqi Yao and Xinyi Ye and Yan Ding and Zhigang Wang and JiaYuan Gu and Bin Zhao and Dong Wang and Xuelong Li},
      year={2025},
      eprint={2501.15830},
      archivePrefix={arXiv},
      primaryClass={cs.RO},
      url={https://arxiv.org/abs/2501.15830}, 
}

@misc{fast,
      title={FAST: Efficient Action Tokenization for Vision-Language-Action Models}, 
      author={Karl Pertsch and Kyle Stachowicz and Brian Ichter and Danny Driess and Suraj Nair and Quan Vuong and Oier Mees and Chelsea Finn and Sergey Levine},
      year={2025},
      eprint={2501.09747},
      archivePrefix={arXiv},
      primaryClass={cs.RO},
      url={https://arxiv.org/abs/2501.09747}, 
}

@article{diffusion_policy,
author = {Bekris, Kostas and Hauser, Kris and Herbert, Sylvia and Yu, Jingjin and Chi, Cheng and Xu, Zhenjia and Feng, Siyuan and Cousineau, Eric and Du, Yilun and Burchfiel, Benjamin and Tedrake, Russ and Song, Shuran},
title = {Diffusion policy: Visuomotor policy learning via action diffusion},
year = {2025},
issue_date = {Sep 2025},
publisher = {Sage Publications, Inc.},
address = {USA},
volume = {44},
number = {10–11},
issn = {0278-3649},
url = {https://doi.org/10.1177/02783649241273668},
doi = {10.1177/02783649241273668},
journal = {Int. J. Rob. Res.},
month = sep,
pages = {1684–1704},
numpages = {21}
}

@misc{rdt1b,
      title={RDT-1B: a Diffusion Foundation Model for Bimanual Manipulation}, 
      author={Songming Liu and Lingxuan Wu and Bangguo Li and Hengkai Tan and Huayu Chen and Zhengyi Wang and Ke Xu and Hang Su and Jun Zhu},
      year={2025},
      eprint={2410.07864},
      archivePrefix={arXiv},
      primaryClass={cs.RO},
      url={https://arxiv.org/abs/2410.07864}, 
}

@misc{hybridvla,
      title={HybridVLA: Collaborative Diffusion and Autoregression in a Unified Vision-Language-Action Model}, 
      author={Jiaming Liu and Hao Chen and Pengju An and Zhuoyang Liu and Renrui Zhang and Chenyang Gu and Xiaoqi Li and Ziyu Guo and Sixiang Chen and Mengzhen Liu and Chengkai Hou and Mengdi Zhao and KC alex Zhou and Pheng-Ann Heng and Shanghang Zhang},
      year={2025},
      eprint={2503.10631},
      archivePrefix={arXiv},
      primaryClass={cs.CV},
      url={https://arxiv.org/abs/2503.10631}, 
}

@misc{gr3,
      title={GR-3 Technical Report}, 
      author={Chilam Cheang and Sijin Chen and Zhongren Cui and Yingdong Hu and Liqun Huang and Tao Kong and Hang Li and Yifeng Li and Yuxiao Liu and Xiao Ma and Hao Niu and Wenxuan Ou and Wanli Peng and Zeyu Ren and Haixin Shi and Jiawen Tian and Hongtao Wu and Xin Xiao and Yuyang Xiao and Jiafeng Xu and Yichu Yang},
      year={2025},
      eprint={2507.15493},
      archivePrefix={arXiv},
      primaryClass={cs.RO},
      url={https://arxiv.org/abs/2507.15493}, 
}

@misc{ddvla,
      title={Discrete Diffusion VLA: Bringing Discrete Diffusion to Action Decoding in Vision-Language-Action Policies}, 
      author={Zhixuan Liang and Yizhuo Li and Tianshuo Yang and Chengyue Wu and Sitong Mao and Tian Nian and Liuao Pei and Shunbo Zhou and Xiaokang Yang and Jiangmiao Pang and Yao Mu and Ping Luo},
      year={2025},
      eprint={2508.20072},
      archivePrefix={arXiv},
      primaryClass={cs.CV},
      url={https://arxiv.org/abs/2508.20072}, 
}

@misc{dvla,
      title={dVLA: Diffusion Vision-Language-Action Model with Multimodal Chain-of-Thought}, 
      author={Junjie Wen and Minjie Zhu and Jiaming Liu and Zhiyuan Liu and Yicun Yang and Linfeng Zhang and Shanghang Zhang and Yichen Zhu and Yi Xu},
      year={2025},
      eprint={2509.25681},
      archivePrefix={arXiv},
      primaryClass={cs.RO},
      url={https://arxiv.org/abs/2509.25681}, 
}

@misc{mmadavla,
      title={MMaDA-VLA: Large Diffusion Vision-Language-Action Model with Unified Multi-Modal Instruction and Generation}, 
      author={Yang Liu and Pengxiang Ding and Tengyue Jiang and Xudong Wang and Wenxuan Song and Minghui Lin and Han Zhao and Hongyin Zhang and Zifeng Zhuang and Wei Zhao and Siteng Huang and Jinkui Shi and Donglin Wang},
      year={2026},
      eprint={2603.25406},
      archivePrefix={arXiv},
      primaryClass={cs.RO},
      url={https://arxiv.org/abs/2603.25406}, 
}

@misc{rt1,
      title={RT-1: Robotics Transformer for Real-World Control at Scale}, 
      author={Anthony Brohan and Noah Brown and Justice Carbajal and Yevgen Chebotar and Joseph Dabis and Chelsea Finn and Keerthana Gopalakrishnan and Karol Hausman and Alex Herzog and Jasmine Hsu and Julian Ibarz and Brian Ichter and Alex Irpan and Tomas Jackson and Sally Jesmonth and Nikhil J Joshi and Ryan Julian and Dmitry Kalashnikov and Yuheng Kuang and Isabel Leal and Kuang-Huei Lee and Sergey Levine and Yao Lu and Utsav Malla and Deeksha Manjunath and Igor Mordatch and Ofir Nachum and Carolina Parada and Jodilyn Peralta and Emily Perez and Karl Pertsch and Jornell Quiambao and Kanishka Rao and Michael Ryoo and Grecia Salazar and Pannag Sanketi and Kevin Sayed and Jaspiar Singh and Sumedh Sontakke and Austin Stone and Clayton Tan and Huong Tran and Vincent Vanhoucke and Steve Vega and Quan Vuong and Fei Xia and Ted Xiao and Peng Xu and Sichun Xu and Tianhe Yu and Brianna Zitkovich},
      year={2023},
      eprint={2212.06817},
      archivePrefix={arXiv},
      primaryClass={cs.RO},
      url={https://arxiv.org/abs/2212.06817}, 
}

@misc{geminirobotics,
      title={Gemini Robotics: Bringing AI into the Physical World}, 
      author={Gemini Robotics Team and Saminda Abeyruwan and Joshua Ainslie and Jean-Baptiste Alayrac and Montserrat Gonzalez Arenas and Travis Armstrong and Ashwin Balakrishna and Robert Baruch and Maria Bauza and Michiel Blokzijl and Steven Bohez and Konstantinos Bousmalis and Anthony Brohan and Thomas Buschmann and Arunkumar Byravan and Serkan Cabi and Ken Caluwaerts and Federico Casarini and Oscar Chang and Jose Enrique Chen and Xi Chen and Hao-Tien Lewis Chiang and Krzysztof Choromanski and David D'Ambrosio and Sudeep Dasari and Todor Davchev and Coline Devin and Norman Di Palo and Tianli Ding and Adil Dostmohamed and Danny Driess and Yilun Du and Debidatta Dwibedi and Michael Elabd and Claudio Fantacci and Cody Fong and Erik Frey and Chuyuan Fu and Marissa Giustina and Keerthana Gopalakrishnan and Laura Graesser and Leonard Hasenclever and Nicolas Heess and Brandon Hernaez and Alexander Herzog and R. Alex Hofer and Jan Humplik and Atil Iscen and Mithun George Jacob and Deepali Jain and Ryan Julian and Dmitry Kalashnikov and M. Emre Karagozler and Stefani Karp and Chase Kew and Jerad Kirkland and Sean Kirmani and Yuheng Kuang and Thomas Lampe and Antoine Laurens and Isabel Leal and Alex X. Lee and Tsang-Wei Edward Lee and Jacky Liang and Yixin Lin and Sharath Maddineni and Anirudha Majumdar and Assaf Hurwitz Michaely and Robert Moreno and Michael Neunert and Francesco Nori and Carolina Parada and Emilio Parisotto and Peter Pastor and Acorn Pooley and Kanishka Rao and Krista Reymann and Dorsa Sadigh and Stefano Saliceti and Pannag Sanketi and Pierre Sermanet and Dhruv Shah and Mohit Sharma and Kathryn Shea and Charles Shu and Vikas Sindhwani and Sumeet Singh and Radu Soricut and Jost Tobias Springenberg and Rachel Sterneck and Razvan Surdulescu and Jie Tan and Jonathan Tompson and Vincent Vanhoucke and Jake Varley and Grace Vesom and Giulia Vezzani and Oriol Vinyals and Ayzaan Wahid and Stefan Welker and Paul Wohlhart and Fei Xia and Ted Xiao and Annie Xie and Jinyu Xie and Peng Xu and Sichun Xu and Ying Xu and Zhuo Xu and Yuxiang Yang and Rui Yao and Sergey Yaroshenko and Wenhao Yu and Wentao Yuan and Jingwei Zhang and Tingnan Zhang and Allan Zhou and Yuxiang Zhou},
      year={2025},
      eprint={2503.20020},
      archivePrefix={arXiv},
      primaryClass={cs.RO},
      url={https://arxiv.org/abs/2503.20020}, 
}

@misc{asyncvla,
      title={AsyncVLA: Asynchronous Flow Matching for Vision-Language-Action Models}, 
      author={Yuhua Jiang and Shuang Cheng and Yan Ding and Feifei Gao and Biqing Qi},
      year={2025},
      eprint={2511.14148},
      archivePrefix={arXiv},
      primaryClass={cs.RO},
      url={https://arxiv.org/abs/2511.14148}, 
}

@misc{udvla,
      title={Unified Diffusion VLA: Vision-Language-Action Model via Joint Discrete Denoising Diffusion Process}, 
      author={Jiayi Chen and Wenxuan Song and Pengxiang Ding and Ziyang Zhou and Han Zhao and Feilong Tang and Donglin Wang and Haoang Li},
      year={2026},
      eprint={2511.01718},
      archivePrefix={arXiv},
      primaryClass={cs.RO},
      url={https://arxiv.org/abs/2511.01718}, 
}

@misc{lladavla,
      title={LLaDA-VLA: Vision Language Diffusion Action Models}, 
      author={Yuqing Wen and Hebei Li and Kefan Gu and Yucheng Zhao and Tiancai Wang and Xiaoyan Sun},
      year={2025},
      eprint={2509.06932},
      archivePrefix={arXiv},
      primaryClass={cs.RO},
      url={https://arxiv.org/abs/2509.06932}, 
}

@misc{a2c2,
      title={Leave No Observation Behind: Real-time Correction for VLA Action Chunks}, 
      author={Kohei Sendai and Maxime Alvarez and Tatsuya Matsushima and Yutaka Matsuo and Yusuke Iwasawa},
      year={2025},
      eprint={2509.23224},
      archivePrefix={arXiv},
      primaryClass={cs.RO},
      url={https://arxiv.org/abs/2509.23224}, 
}

@misc{autohorizon,
      title={VLA Knows Its Limits}, 
      author={Haoxuan Wang and Gengyu Zhang and Yan Yan and Ramana Rao Kompella and Gaowen Liu},
      year={2026},
      eprint={2602.21445},
      archivePrefix={arXiv},
      primaryClass={cs.RO},
      url={https://arxiv.org/abs/2602.21445}, 
}

@misc{pi0,
      title={$\pi_0$: A Vision-Language-Action Flow Model for General Robot Control}, 
      author={Kevin Black and Noah Brown and Danny Driess and Adnan Esmail and Michael Equi and Chelsea Finn and Niccolo Fusai and Lachy Groom and Karol Hausman and Brian Ichter and Szymon Jakubczak and Tim Jones and Liyiming Ke and Sergey Levine and Adrian Li-Bell and Mohith Mothukuri and Suraj Nair and Karl Pertsch and Lucy Xiaoyang Shi and James Tanner and Quan Vuong and Anna Walling and Haohuan Wang and Ury Zhilinsky},
      year={2026},
      eprint={2410.24164},
      archivePrefix={arXiv},
      primaryClass={cs.LG},
      url={https://arxiv.org/abs/2410.24164}, 
}

@misc{rt2,
      title={RT-2: Vision-Language-Action Models Transfer Web Knowledge to Robotic Control}, 
      author={Anthony Brohan and Noah Brown and Justice Carbajal and Yevgen Chebotar and Xi Chen and Krzysztof Choromanski and Tianli Ding and Danny Driess and Avinava Dubey and Chelsea Finn and Pete Florence and Chuyuan Fu and Montse Gonzalez Arenas and Keerthana Gopalakrishnan and Kehang Han and Karol Hausman and Alexander Herzog and Jasmine Hsu and Brian Ichter and Alex Irpan and Nikhil Joshi and Ryan Julian and Dmitry Kalashnikov and Yuheng Kuang and Isabel Leal and Lisa Lee and Tsang-Wei Edward Lee and Sergey Levine and Yao Lu and Henryk Michalewski and Igor Mordatch and Karl Pertsch and Kanishka Rao and Krista Reymann and Michael Ryoo and Grecia Salazar and Pannag Sanketi and Pierre Sermanet and Jaspiar Singh and Anikait Singh and Radu Soricut and Huong Tran and Vincent Vanhoucke and Quan Vuong and Ayzaan Wahid and Stefan Welker and Paul Wohlhart and Jialin Wu and Fei Xia and Ted Xiao and Peng Xu and Sichun Xu and Tianhe Yu and Brianna Zitkovich},
      year={2023},
      eprint={2307.15818},
      archivePrefix={arXiv},
      primaryClass={cs.RO},
      url={https://arxiv.org/abs/2307.15818}, 
}

@misc{qwenvl,
      title={Qwen-VL: A Versatile Vision-Language Model for Understanding, Localization, Text Reading, and Beyond}, 
      author={Jinze Bai and Shuai Bai and Shusheng Yang and Shijie Wang and Sinan Tan and Peng Wang and Junyang Lin and Chang Zhou and Jingren Zhou},
      year={2023},
      eprint={2308.12966},
      archivePrefix={arXiv},
      primaryClass={cs.CV},
      url={https://arxiv.org/abs/2308.12966}, 
}

@misc{llama2,
      title={Llama 2: Open Foundation and Fine-Tuned Chat Models}, 
      author={Hugo Touvron and Louis Martin and Kevin Stone and Peter Albert and Amjad Almahairi and Yasmine Babaei and Nikolay Bashlykov and Soumya Batra and Prajjwal Bhargava and Shruti Bhosale and Dan Bikel and Lukas Blecher and Cristian Canton Ferrer and Moya Chen and Guillem Cucurull and David Esiobu and Jude Fernandes and Jeremy Fu and Wenyin Fu and Brian Fuller and Cynthia Gao and Vedanuj Goswami and Naman Goyal and Anthony Hartshorn and Saghar Hosseini and Rui Hou and Hakan Inan and Marcin Kardas and Viktor Kerkez and Madian Khabsa and Isabel Kloumann and Artem Korenev and Punit Singh Koura and Marie-Anne Lachaux and Thibaut Lavril and Jenya Lee and Diana Liskovich and Yinghai Lu and Yuning Mao and Xavier Martinet and Todor Mihaylov and Pushkar Mishra and Igor Molybog and Yixin Nie and Andrew Poulton and Jeremy Reizenstein and Rashi Rungta and Kalyan Saladi and Alan Schelten and Ruan Silva and Eric Michael Smith and Ranjan Subramanian and Xiaoqing Ellen Tan and Binh Tang and Ross Taylor and Adina Williams and Jian Xiang Kuan and Puxin Xu and Zheng Yan and Iliyan Zarov and Yuchen Zhang and Angela Fan and Melanie Kambadur and Sharan Narang and Aurelien Rodriguez and Robert Stojnic and Sergey Edunov and Thomas Scialom},
      year={2023},
      eprint={2307.09288},
      archivePrefix={arXiv},
      primaryClass={cs.CL},
      url={https://arxiv.org/abs/2307.09288}, 
}

@misc{openx,
title={Open {X-E}mbodiment: Robotic Learning Datasets and {RT-X} Models},
author = {Open X-Embodiment Collaboration and Abby O'Neill and Abdul Rehman and Abhinav Gupta and Abhiram Maddukuri and Abhishek Gupta and Abhishek Padalkar and Abraham Lee and Acorn Pooley and Agrim Gupta and Ajay Mandlekar and Ajinkya Jain and Albert Tung and Alex Bewley and Alex Herzog and Alex Irpan and Alexander Khazatsky and Anant Rai and Anchit Gupta and Andrew Wang and Andrey Kolobov and Anikait Singh and Animesh Garg and Aniruddha Kembhavi and Annie Xie and Anthony Brohan and Antonin Raffin and Archit Sharma and Arefeh Yavary and Arhan Jain and Ashwin Balakrishna and Ayzaan Wahid and Ben Burgess-Limerick and Beomjoon Kim and Bernhard Schölkopf and Blake Wulfe and Brian Ichter and Cewu Lu and Charles Xu and Charlotte Le and Chelsea Finn and Chen Wang and Chenfeng Xu and Cheng Chi and Chenguang Huang and Christine Chan and Christopher Agia and Chuer Pan and Chuyuan Fu and Coline Devin and Danfei Xu and Daniel Morton and Danny Driess and Daphne Chen and Deepak Pathak and Dhruv Shah and Dieter Büchler and Dinesh Jayaraman and Dmitry Kalashnikov and Dorsa Sadigh and Edward Johns and Ethan Foster and Fangchen Liu and Federico Ceola and Fei Xia and Feiyu Zhao and Felipe Vieira Frujeri and Freek Stulp and Gaoyue Zhou and Gaurav S. Sukhatme and Gautam Salhotra and Ge Yan and Gilbert Feng and Giulio Schiavi and Glen Berseth and Gregory Kahn and Guangwen Yang and Guanzhi Wang and Hao Su and Hao-Shu Fang and Haochen Shi and Henghui Bao and Heni Ben Amor and Henrik I Christensen and Hiroki Furuta and Homanga Bharadhwaj and Homer Walke and Hongjie Fang and Huy Ha and Igor Mordatch and Ilija Radosavovic and Isabel Leal and Jacky Liang and Jad Abou-Chakra and Jaehyung Kim and Jaimyn Drake and Jan Peters and Jan Schneider and Jasmine Hsu and Jay Vakil and Jeannette Bohg and Jeffrey Bingham and Jeffrey Wu and Jensen Gao and Jiaheng Hu and Jiajun Wu and Jialin Wu and Jiankai Sun and Jianlan Luo and Jiayuan Gu and Jie Tan and Jihoon Oh and Jimmy Wu and Jingpei Lu and Jingyun Yang and Jitendra Malik and João Silvério and Joey Hejna and Jonathan Booher and Jonathan Tompson and Jonathan Yang and Jordi Salvador and Joseph J. Lim and Junhyek Han and Kaiyuan Wang and Kanishka Rao and Karl Pertsch and Karol Hausman and Keegan Go and Keerthana Gopalakrishnan and Ken Goldberg and Kendra Byrne and Kenneth Oslund and Kento Kawaharazuka and Kevin Black and Kevin Lin and Kevin Zhang and Kiana Ehsani and Kiran Lekkala and Kirsty Ellis and Krishan Rana and Krishnan Srinivasan and Kuan Fang and Kunal Pratap Singh and Kuo-Hao Zeng and Kyle Hatch and Kyle Hsu and Laurent Itti and Lawrence Yunliang Chen and Lerrel Pinto and Li Fei-Fei and Liam Tan and Linxi "Jim" Fan and Lionel Ott and Lisa Lee and Luca Weihs and Magnum Chen and Marion Lepert and Marius Memmel and Masayoshi Tomizuka and Masha Itkina and Mateo Guaman Castro and Max Spero and Maximilian Du and Michael Ahn and Michael C. Yip and Mingtong Zhang and Mingyu Ding and Minho Heo and Mohan Kumar Srirama and Mohit Sharma and Moo Jin Kim and Muhammad Zubair Irshad and Naoaki Kanazawa and Nicklas Hansen and Nicolas Heess and Nikhil J Joshi and Niko Suenderhauf and Ning Liu and Norman Di Palo and Nur Muhammad Mahi Shafiullah and Oier Mees and Oliver Kroemer and Osbert Bastani and Pannag R Sanketi and Patrick "Tree" Miller and Patrick Yin and Paul Wohlhart and Peng Xu and Peter David Fagan and Peter Mitrano and Pierre Sermanet and Pieter Abbeel and Priya Sundaresan and Qiuyu Chen and Quan Vuong and Rafael Rafailov and Ran Tian and Ria Doshi and Roberto Mart{'i}n-Mart{'i}n and Rohan Baijal and Rosario Scalise and Rose Hendrix and Roy Lin and Runjia Qian and Ruohan Zhang and Russell Mendonca and Rutav Shah and Ryan Hoque and Ryan Julian and Samuel Bustamante and Sean Kirmani and Sergey Levine and Shan Lin and Sherry Moore and Shikhar Bahl and Shivin Dass and Shubham Sonawani and Shubham Tulsiani and Shuran Song and Sichun Xu and Siddhant Haldar and Siddharth Karamcheti and Simeon Adebola and Simon Guist and Soroush Nasiriany and Stefan Schaal and Stefan Welker and Stephen Tian and Subramanian Ramamoorthy and Sudeep Dasari and Suneel Belkhale and Sungjae Park and Suraj Nair and Suvir Mirchandani and Takayuki Osa and Tanmay Gupta and Tatsuya Harada and Tatsuya Matsushima and Ted Xiao and Thomas Kollar and Tianhe Yu and Tianli Ding and Todor Davchev and Tony Z. Zhao and Travis Armstrong and Trevor Darrell and Trinity Chung and Vidhi Jain and Vikash Kumar and Vincent Vanhoucke and Vitor Guizilini and Wei Zhan and Wenxuan Zhou and Wolfram Burgard and Xi Chen and Xiangyu Chen and Xiaolong Wang and Xinghao Zhu and Xinyang Geng and Xiyuan Liu and Xu Liangwei and Xuanlin Li and Yansong Pang and Yao Lu and Yecheng Jason Ma and Yejin Kim and Yevgen Chebotar and Yifan Zhou and Yifeng Zhu and Yilin Wu and Ying Xu and Yixuan Wang and Yonatan Bisk and Yongqiang Dou and Yoonyoung Cho and Youngwoon Lee and Yuchen Cui and Yue Cao and Yueh-Hua Wu and Yujin Tang and Yuke Zhu and Yunchu Zhang and Yunfan Jiang and Yunshuang Li and Yunzhu Li and Yusuke Iwasawa and Yutaka Matsuo and Zehan Ma and Zhuo Xu and Zichen Jeff Cui and Zichen Zhang and Zipeng Fu and Zipeng Lin},
howpublished  = {\url{https://arxiv.org/abs/2310.08864}},
year = {2023},
}

@misc{bridge,
      title={BridgeData V2: A Dataset for Robot Learning at Scale}, 
      author={Homer Walke and Kevin Black and Abraham Lee and Moo Jin Kim and Max Du and Chongyi Zheng and Tony Zhao and Philippe Hansen-Estruch and Quan Vuong and Andre He and Vivek Myers and Kuan Fang and Chelsea Finn and Sergey Levine},
      year={2024},
      eprint={2308.12952},
      archivePrefix={arXiv},
      primaryClass={cs.RO},
      url={https://arxiv.org/abs/2308.12952}, 
}

@article{gpucke,
author = {Lin, Zhen and Dai, Hongwen and Mantor, Michael and Zhou, Huiyang},
title = {Coordinated CTA Combination and Bandwidth Partitioning for GPU Concurrent Kernel Execution},
year = {2019},
issue_date = {September 2019},
publisher = {Association for Computing Machinery},
address = {New York, NY, USA},
volume = {16},
number = {3},
issn = {1544-3566},
url = {https://doi.org/10.1145/3326124},
doi = {10.1145/3326124},
journal = {ACM Trans. Archit. Code Optim.},
month = jun,
articleno = {23},
numpages = {27}
}

@ARTICLE{kim2019dynsharing,
author={Kim, Jiho and Cha, Jehee and Park, Jason Jong Kyu and Jeon, Dongsuk and Park, Yongjun},
journal={ IEEE Computer Architecture Letters },
title={{ Improving GPU Multitasking Efficiency Using Dynamic Resource Sharing }},
year={2019},
volume={18},
number={01},
ISSN={1556-6064},
pages={1-5},
doi={10.1109/LCA.2018.2889042},
url = {https://doi.ieeecomputersociety.org/10.1109/LCA.2018.2889042},
publisher={IEEE Computer Society},
address={Los Alamitos, CA, USA},
month=jan}

@article{kim2021kscheduler,
author = {Kim, Sejin and Kim, Yoonhee},
title = {K-Scheduler: dynamic intra-SM multitasking management with execution profiles on GPUs},
year = {2022},
issue_date = {Feb 2022},
publisher = {Kluwer Academic Publishers},
address = {USA},
volume = {25},
number = {1},
issn = {1386-7857},
url = {https://doi.org/10.1007/s10586-021-03429-7},
doi = {10.1007/s10586-021-03429-7},
journal = {Cluster Computing},
month = feb,
pages = {597–617},
numpages = {21}
}

@article{liang2020modelbasedsmk,
author = {Wu, Hao and Liu, Weizhi and Lin, Huanxin and Wang, Cho-Li},
title = {A Model-Based Software Solution for Simultaneous Multiple Kernels on GPUs},
year = {2020},
issue_date = {March 2020},
publisher = {Association for Computing Machinery},
address = {New York, NY, USA},
volume = {17},
number = {1},
issn = {1544-3566},
url = {https://doi.org/10.1145/3377138},
doi = {10.1145/3377138},
journal = {ACM Trans. Archit. Code Optim.},
month = mar,
articleno = {7},
numpages = {26}
}

@INPROCEEDINGS{zhao2021plasticine,
  author={Zhao, Han and Cui, Weihao and Chen, Quan and Zhao, Jieru and Leng, Jingwen and Guo, Minyi},
  booktitle={2021 IEEE 39th International Conference on Computer Design (ICCD)}, 
  title={Exploiting Intra-SM Parallelism in GPUs via Persistent and Elastic Blocks}, 
  year={2021},
  volume={},
  number={},
  pages={290-298},
  doi={10.1109/ICCD53106.2021.00054}}

@article{pai2013elastic,
author = {Pai, Sreepathi and Thazhuthaveetil, Matthew J. and Govindarajan, R.},
title = {Improving GPGPU concurrency with elastic kernels},
year = {2013},
issue_date = {March 2013},
publisher = {Association for Computing Machinery},
address = {New York, NY, USA},
volume = {41},
number = {1},
issn = {0163-5964},
url = {https://doi.org/10.1145/2490301.2451160},
doi = {10.1145/2490301.2451160},
journal = {SIGARCH Comput. Archit. News},
month = mar,
pages = {407–418},
numpages = {12}
}

@INPROCEEDINGS{allen2019slate,
  author={Allen, Tyler and Feng, Xizhou and Ge, Rong},
  booktitle={2019 IEEE International Parallel and Distributed Processing Symposium (IPDPS)}, 
  title={Slate: Enabling Workload-Aware Efficient Multiprocessing for Modern GPGPUs}, 
  year={2019},
  volume={},
  number={},
  pages={252-261},
  doi={10.1109/IPDPS.2019.00035}}

@ARTICLE{shekofteh2020ccuda,
author={Shekofteh, S.-Kazem and Noori, Hamid and Naghibzadeh, Mahmoud and Froning, Holger and Yazdi, Hadi Sadoghi},
journal={ IEEE Transactions on Parallel \& Distributed Systems },
title={{ cCUDA: Effective Co-Scheduling of Concurrent Kernels on GPUs }},
year={2020},
volume={31},
number={04},
ISSN={1558-2183},
pages={766-778},
doi={10.1109/TPDS.2019.2944602},
url = {https://doi.ieeecomputersociety.org/10.1109/TPDS.2019.2944602},
publisher={IEEE Computer Society},
address={Los Alamitos, CA, USA},
month=apr}

@inproceedings{chen2016baymax,
author = {Chen, Quan and Yang, Hailong and Mars, Jason and Tang, Lingjia},
title = {Baymax: QoS Awareness and Increased Utilization for Non-Preemptive Accelerators in Warehouse Scale Computers},
year = {2016},
isbn = {9781450340915},
publisher = {Association for Computing Machinery},
address = {New York, NY, USA},
url = {https://doi.org/10.1145/2872362.2872368},
doi = {10.1145/2872362.2872368},
booktitle = {Proceedings of the Twenty-First International Conference on Architectural Support for Programming Languages and Operating Systems},
pages = {681–696},
numpages = {16},
location = {Atlanta, Georgia, USA},
series = {ASPLOS '16}
}

@article{chen2017prophet,
author = {Chen, Quan and Yang, Hailong and Guo, Minyi and Kannan, Ram Srivatsa and Mars, Jason and Tang, Lingjia},
title = {Prophet: Precise QoS Prediction on Non-Preemptive Accelerators to Improve Utilization in Warehouse-Scale Computers},
year = {2017},
issue_date = {March 2017},
publisher = {Association for Computing Machinery},
address = {New York, NY, USA},
volume = {45},
number = {1},
issn = {0163-5964},
url = {https://doi.org/10.1145/3093337.3037700},
doi = {10.1145/3093337.3037700},
journal = {SIGARCH Comput. Archit. News},
month = apr,
pages = {17–32},
numpages = {16}
}

@INPROCEEDINGS {li2022hfuse,
author = { Li, Ao and Zheng, Bojian and Pekhimenko, Gennady and Long, Fan },
booktitle = { 2022 IEEE/ACM International Symposium on Code Generation and Optimization (CGO) },
title = {{ Automatic Horizontal Fusion for GPU Kernels }},
year = {2022},
volume = {},
ISSN = {},
pages = {14-27},
doi = {10.1109/CGO53902.2022.9741270},
url = {https://doi.ieeecomputersociety.org/10.1109/CGO53902.2022.9741270},
publisher = {IEEE Computer Society},
address = {Los Alamitos, CA, USA},
month =apr}

@article{park2015chimera,
author = {Park, Jason Jong Kyu and Park, Yongjun and Mahlke, Scott},
title = {Chimera: Collaborative Preemption for Multitasking on a Shared GPU},
year = {2015},
issue_date = {April 2015},
publisher = {Association for Computing Machinery},
address = {New York, NY, USA},
volume = {50},
number = {4},
issn = {0362-1340},
url = {https://doi.org/10.1145/2775054.2694346},
doi = {10.1145/2775054.2694346},
journal = {SIGPLAN Not.},
month = mar,
pages = {593–606},
numpages = {14}
}

@INPROCEEDINGS {tanasic2014srtf,
author = { Pai, Sreepathi and Govindarajan, R. and Thazhuthaveetil, Matthew J. },
booktitle = { 2014 23rd International Conference on Parallel Architecture and Compilation (PACT) },
title = {{ Preemptive thread block scheduling with online structural runtime prediction for concurrent GPGPU kernels }},
year = {2014},
volume = {},
ISSN = {},
pages = {483-484},
doi = {10.1145/2628071.2628117},
url = {https://doi.ieeecomputersociety.org/10.1145/2628071.2628117},
publisher = {IEEE Computer Society},
address = {Los Alamitos, CA, USA},
month =Aug}

@inproceedings{wu2017flep,
author = {Wu, Bo and Liu, Xu and Zhou, Xiaobo and Jiang, Changjun},
title = {FLEP: Enabling Flexible and Efficient Preemption on GPUs},
year = {2017},
isbn = {9781450344654},
publisher = {Association for Computing Machinery},
address = {New York, NY, USA},
url = {https://doi.org/10.1145/3037697.3037742},
doi = {10.1145/3037697.3037742},
booktitle = {Proceedings of the Twenty-Second International Conference on Architectural Support for Programming Languages and Operating Systems},
pages = {483–496},
numpages = {14},
location = {Xi'an, China},
series = {ASPLOS '17}
}

@inproceedings{kayiran2016ucstates,
author = {Kayiran, Onur and Jog, Adwait and Pattnaik, Ashutosh and Ausavarungnirun, Rachata and Tang, Xulong and Kandemir, Mahmut T. and Loh, Gabriel H. and Mutlu, Onur and Das, Chita R.},
title = {$\mu$C-States: Fine-grained GPU Datapath Power Management},
year = {2016},
isbn = {9781450341219},
publisher = {Association for Computing Machinery},
address = {New York, NY, USA},
url = {https://doi.org/10.1145/2967938.2967941},
doi = {10.1145/2967938.2967941},
booktitle = {Proceedings of the 2016 International Conference on Parallel Architectures and Compilation},
pages = {17–30},
numpages = {14},
location = {Haifa, Israel},
series = {PACT '16}
}

@inproceedings{wu2015smcentric,
author = {Wu, Bo and Chen, Guoyang and Li, Dong and Shen, Xipeng and Vetter, Jeffrey},
title = {Enabling and Exploiting Flexible Task Assignment on GPU through SM-Centric Program Transformations},
year = {2015},
isbn = {9781450335591},
publisher = {Association for Computing Machinery},
address = {New York, NY, USA},
url = {https://doi.org/10.1145/2751205.2751213},
doi = {10.1145/2751205.2751213},
booktitle = {Proceedings of the 29th ACM on International Conference on Supercomputing},
pages = {119–130},
numpages = {12},
location = {Newport Beach, California, USA},
series = {ICS '15}
}

@INPROCEEDINGS{narasiman2011lwm,
  author={Narasiman, Veynu and Shebanow, Michael and Lee, Chang Joo and Miftakhutdinov, Rustam and Mutlu, Onur and Patt, Yale N.},
  booktitle={2011 44th Annual IEEE/ACM International Symposium on Microarchitecture (MICRO)}, 
  title={Improving GPU performance via large warps and two-level warp scheduling}, 
  year={2011},
  volume={},
  number={},
  pages={308-317},
  doi={}}

@inproceedings{rogers2012ccws,
author = {Rogers, Timothy G. and O'Connor, Mike and Aamodt, Tor M.},
title = {Cache-Conscious Wavefront Scheduling},
year = {2012},
isbn = {9780769549248},
publisher = {IEEE Computer Society},
address = {USA},
url = {https://doi.org/10.1109/MICRO.2012.16},
doi = {10.1109/MICRO.2012.16},
booktitle = {Proceedings of the 2012 45th Annual IEEE/ACM International Symposium on Microarchitecture},
pages = {72–83},
numpages = {12},
location = {Vancouver, B.C., CANADA},
series = {MICRO-45}
}

@inproceedings{rogers2013daws,
author = {Rogers, Timothy G. and O'Connor, Mike and Aamodt, Tor M.},
title = {Divergence-aware warp scheduling},
year = {2013},
isbn = {9781450326384},
publisher = {Association for Computing Machinery},
address = {New York, NY, USA},
url = {https://doi.org/10.1145/2540708.2540718},
doi = {10.1145/2540708.2540718},
booktitle = {Proceedings of the 46th Annual IEEE/ACM International Symposium on Microarchitecture},
pages = {99–110},
numpages = {12},
location = {Davis, California},
series = {MICRO-46}
}

@inproceedings{lee2014caws,
author = {Lee, Shin-Ying and Wu, Carole-Jean},
title = {CAWS: criticality-aware warp scheduling for GPGPU workloads},
year = {2014},
isbn = {9781450328098},
publisher = {Association for Computing Machinery},
address = {New York, NY, USA},
url = {https://doi.org/10.1145/2628071.2628107},
doi = {10.1145/2628071.2628107},
booktitle = {Proceedings of the 23rd International Conference on Parallel Architectures and Compilation},
pages = {175–186},
numpages = {12},
location = {Edmonton, AB, Canada},
series = {PACT '14}
}

@article{lin2019coordinated,
author = {Lin, Zhen and Dai, Hongwen and Mantor, Michael and Zhou, Huiyang},
title = {Coordinated CTA Combination and Bandwidth Partitioning for GPU Concurrent Kernel Execution},
year = {2019},
issue_date = {September 2019},
publisher = {Association for Computing Machinery},
address = {New York, NY, USA},
volume = {16},
number = {3},
issn = {1544-3566},
url = {https://doi.org/10.1145/3326124},
doi = {10.1145/3326124},
journal = {ACM Trans. Archit. Code Optim.},
month = jun,
articleno = {23},
numpages = {27}
}

@article{flipsteal,
author = {Huzaifa, Muhammad and Alsop, Johnathan and Mahmoud, Abdulrahman and Salvador, Giordano and Sinclair, Matthew D. and Adve, Sarita V.},
title = {Inter-kernel Reuse-aware Thread Block Scheduling},
year = {2020},
issue_date = {September 2020},
publisher = {Association for Computing Machinery},
address = {New York, NY, USA},
volume = {17},
number = {3},
issn = {1544-3566},
url = {https://doi.org/10.1145/3406538},
doi = {10.1145/3406538},
journal = {ACM Trans. Archit. Code Optim.},
month = aug,
articleno = {24},
numpages = {27}
}

@article{paver,
author = {Tripathy, Devashree and Abdolrashidi, Amirali and Bhuyan, Laxmi Narayan and Zhou, Liang and Wong, Daniel},
title = {PAVER: Locality Graph-Based Thread Block Scheduling for GPUs},
year = {2021},
issue_date = {September 2021},
publisher = {Association for Computing Machinery},
address = {New York, NY, USA},
volume = {18},
number = {3},
issn = {1544-3566},
url = {https://doi.org/10.1145/3451164},
doi = {10.1145/3451164},
journal = {ACM Trans. Archit. Code Optim.},
month = jun,
articleno = {32},
numpages = {26}
}

@article{cluster,
author = {Ukarande, Aditya and Patidar, Suryakant and Rangan, Ram},
title = {Locality-Aware CTA Scheduling for Gaming Applications},
year = {2021},
issue_date = {March 2022},
publisher = {Association for Computing Machinery},
address = {New York, NY, USA},
volume = {19},
number = {1},
issn = {1544-3566},
url = {https://doi.org/10.1145/3477497},
doi = {10.1145/3477497},
journal = {ACM Trans. Archit. Code Optim.},
month = dec,
articleno = {1},
numpages = {26}
}

@inproceedings{vijaykumar2015caba,
author = {Vijaykumar, Nandita and Pekhimenko, Gennady and Jog, Adwait and Bhowmick, Abhishek and Ausavarungnirun, Rachata and Das, Chita and Kandemir, Mahmut and Mowry, Todd C. and Mutlu, Onur},
title = {A case for core-assisted bottleneck acceleration in GPUs: enabling flexible data compression with assist warps},
year = {2015},
isbn = {9781450334020},
publisher = {Association for Computing Machinery},
address = {New York, NY, USA},
url = {https://doi.org/10.1145/2749469.2750399},
doi = {10.1145/2749469.2750399},
booktitle = {Proceedings of the 42nd Annual International Symposium on Computer Architecture},
pages = {41–53},
numpages = {13},
location = {Portland, Oregon},
series = {ISCA '15}
}

@inproceedings{latoa,
author = {Bitalebi, Hossein and Geraeinejad, Vahid and Safaei, Farshad and Ebrahimi, Masoumeh},
title = {LATOA: Load-Aware Task Offloading and Adoption in GPU},
year = {2023},
isbn = {9798400707766},
publisher = {Association for Computing Machinery},
address = {New York, NY, USA},
url = {https://doi.org/10.1145/3589236.3589243},
doi = {10.1145/3589236.3589243},
booktitle = {Proceedings of the 15th Workshop on General Purpose Processing Using GPU},
pages = {7–13},
numpages = {7},
location = {Montreal, Canada},
series = {GPGPU '23}
}

@INPROCEEDINGS{awatramani2013kits,
  author={Awatramani, Mihir and Zambreno, Joseph and Rover, Diane},
  booktitle={2013 IEEE 31st International Conference on Computer Design (ICCD)}, 
  title={Increasing GPU throughput using kernel interleaved thread block scheduling}, 
  year={2013},
  volume={},
  number={},
  pages={503-506},
  doi={10.1109/ICCD.2013.6657093}}

@INPROCEEDINGS{adriaens2012spatial,
  author={Adriaens, Jacob T. and Compton, Katherine and Kim, Nam Sung and Schulte, Michael J.},
  booktitle={IEEE International Symposium on High-Performance Comp Architecture}, 
  title={The case for GPGPU spatial multitasking}, 
  year={2012},
  volume={},
  number={},
  pages={1-12},
  doi={10.1109/HPCA.2012.6168946}}

@INPROCEEDINGS{aguilera2014fairshare,
  author={Aguilera, Paula and Morrow, Katherine and Kim, Nam Sung},
  booktitle={2014 IEEE 32nd International Conference on Computer Design (ICCD)}, 
  title={Fair share: Allocation of GPU resources for both performance and fairness}, 
  year={2014},
  volume={},
  number={},
  pages={440-447},
  doi={10.1109/ICCD.2014.6974717}}

@INPROCEEDINGS{aguilera2014qos,
  author={Aguilera, Paula and Morrow, Katherine and Kim, Nam Sung},
  booktitle={2014 19th Asia and South Pacific Design Automation Conference (ASP-DAC)}, 
  title={QoS-aware dynamic resource allocation for spatial-multitasking GPUs}, 
  year={2014},
  volume={},
  number={},
  pages={726-731},
  doi={10.1109/ASPDAC.2014.6742976}}

@article{liang2016kernelmgmt,
author = {Liang, Yun and Li, Xiuhong},
title = {Efficient Kernel Management on GPUs},
year = {2017},
issue_date = {November 2017},
publisher = {Association for Computing Machinery},
address = {New York, NY, USA},
volume = {16},
number = {4},
issn = {1539-9087},
url = {https://doi.org/10.1145/3070710},
doi = {10.1145/3070710},
journal = {ACM Trans. Embed. Comput. Syst.},
month = may,
articleno = {115},
numpages = {24}
}

@inproceedings{zhao2018classification,
author = {Zhao, Xia and Wang, Zhiying and Eeckhout, Lieven},
title = {Classification-Driven Search for Effective SM Partitioning in Multitasking GPUs},
year = {2018},
isbn = {9781450357838},
publisher = {Association for Computing Machinery},
address = {New York, NY, USA},
url = {https://doi.org/10.1145/3205289.3205311},
doi = {10.1145/3205289.3205311},
booktitle = {Proceedings of the 2018 International Conference on Supercomputing},
pages = {65–75},
numpages = {11},
location = {Beijing, China},
series = {ICS '18}
}

@inproceedings{liu2015saws,
author = {Liu, Jiwei and Yang, Jun and Melhem, Rami},
title = {SAWS: synchronization aware GPGPU warp scheduling for multiple independent warp schedulers},
year = {2015},
isbn = {9781450340342},
publisher = {Association for Computing Machinery},
address = {New York, NY, USA},
url = {https://doi.org/10.1145/2830772.2830822},
doi = {10.1145/2830772.2830822},
booktitle = {Proceedings of the 48th International Symposium on Microarchitecture},
pages = {383–394},
numpages = {12},
location = {Waikiki, Hawaii},
series = {MICRO-48}
}

@inproceedings{wang2016laperm,
author = {Wang, Jin and Rubin, Norm and Sidelnik, Albert and Yalamanchili, Sudhakar},
title = {LaPerm: locality aware scheduler for dynamic parallelism on GPUs},
year = {2016},
isbn = {9781467389471},
publisher = {IEEE Press},
url = {https://doi.org/10.1109/ISCA.2016.57},
doi = {10.1109/ISCA.2016.57},
booktitle = {Proceedings of the 43rd International Symposium on Computer Architecture},
pages = {583–595},
numpages = {13},
location = {Seoul, Republic of Korea},
series = {ISCA '16}
}

@ARTICLE{chen2017tbs,
  author={Chen, Li-Jhan and Cheng, Hsiang-Yun and Wang, Po-Han and Yang, Chia-Lin},
  journal={IEEE Computer Architecture Letters}, 
  title={Improving GPGPU Performance via Cache Locality Aware Thread Block Scheduling}, 
  year={2017},
  volume={16},
  number={2},
  pages={127-131},
  doi={10.1109/LCA.2017.2693371}}

@inproceedings{belviranli2016cumas,
author = {Belviranli, Mehmet E. and Khorasani, Farzad and Bhuyan, Laxmi N. and Gupta, Rajiv},
title = {CuMAS: Data Transfer Aware Multi-Application Scheduling for Shared GPUs},
year = {2016},
isbn = {9781450343619},
publisher = {Association for Computing Machinery},
address = {New York, NY, USA},
url = {https://doi.org/10.1145/2925426.2926271},
doi = {10.1145/2925426.2926271},
booktitle = {Proceedings of the 2016 International Conference on Supercomputing},
articleno = {31},
numpages = {12},
location = {Istanbul, Turkey},
series = {ICS '16}
}

@inproceedings{blockmaestro,
author = {Abdolrashidi, AmirAli and Esfeden, Hodjat Asghari and Jahanshahi, Ali and Singh, Kaustubh and Abu-Ghazaleh, Nael and Wong, Daniel},
title = {BlockMaestro: enabling programmer-transparent task-based execution in GPU systems},
year = {2021},
isbn = {9781450390866},
publisher = {IEEE Press},
url = {https://doi.org/10.1109/ISCA52012.2021.00034},
doi = {10.1109/ISCA52012.2021.00034},
booktitle = {Proceedings of the 48th Annual International Symposium on Computer Architecture},
pages = {333–346},
numpages = {14},
location = {Virtual Event, Spain},
series = {ISCA '21}
}

@inproceedings{wireframe,
author = {Abdolrashidi, AmirAli and Tripathy, Devashree and Belviranli, Mehmet Esat and Bhuyan, Laxmi Narayan and Wong, Daniel},
title = {Wireframe: supporting data-dependent parallelism through dependency graph execution in GPUs},
year = {2017},
isbn = {9781450349529},
publisher = {Association for Computing Machinery},
address = {New York, NY, USA},
url = {https://doi.org/10.1145/3123939.3123976},
doi = {10.1145/3123939.3123976},
booktitle = {Proceedings of the 50th Annual IEEE/ACM International Symposium on Microarchitecture},
pages = {600–611},
numpages = {12},
location = {Cambridge, Massachusetts},
series = {MICRO-50 '17}
}










\end{document}